\documentclass[12pt]{iopart}
\eqnobysec
\usepackage{graphicx}

\font\amsb=msbm10
\def\hbar{\mbox{\amsb\char'175}}

\newcommand{\be}{\begin{equation}}
\newcommand{\ee}{\end{equation}}
\newcommand{\vxi}{{\mathbf \xi}}
\newcommand{\veta}{{\mathbf \eta}}
\newcommand{\x}{{\mathbf x}}

\newcommand{\y}{{\mathbf y}}

\newcommand{\vl}{{\mathbf l}}

\newcommand{\J}{{\mathbf J}}

\newcommand{\C}{{\mathbf C}}

\newcommand{\opA}{{\widehat{A}}}

\newcommand{\opR}{{\widehat{R}}}
\newcommand{\opT}{{\widehat{T}}}
\newcommand{\oprho}{{\widehat{\rho}}}

\newcommand{\opH}{\widehat{H}}
\newcommand{\opU}{\widehat{U}}
\newcommand{\opI}{\widehat{I}}

\newcommand{\chf}{\chi}

\newcommand{\der}{\partial}

\newcommand{\GO}{{\mathcal O}}

\begin{document}

\title{Entanglement in phase space}

\author {A. M. Ozorio de Almeida\footnote{ozorio@cbpf.br}}
\address{Centro Brasileiro de Pesquisas Fisicas, 
Rua Xavier Sigaud 150, 22290-180, 
Rio de Janeiro, R.J., Brazil.}

\begin{abstract}

The peculiar effects of a quantum measurement are completely foreign to
classical physics, but relevant features of an entangled quantum
state are accessible to semiclassical analysis.
Classical surfaces, or manifolds in phase space
correspond to quantum states in Hilbert space. 
Subsystems specifie factor spaces of the Hilbert space, 
which correspond to lower dimensional phase spaces.
An entangled state corresponds semiclassically to a surface 
that cannot be decomposed into a product of lower dimensional surfaces,
for a specific choice of subspaces. 
Such a classical factorization never exists for ergodic eigenstates 
of a chaotic Hamiltonian.

Entanglement is best described in the language of density operators.  
The {\it product} of the {\it bra} space and the {\it ket} space
forms a dyadic basis for quantum operators,
analogous to combining the Hilbert spaces for subsystems. 
Thus, the space of quantum operators corresponds to a double phase space. 
The various representations
of the density operator then result from alternative choices
of allowed coordinate planes in this enlarged phase space.
The chosen plane may be a phase space on its own, as in the case of
the Wigner function, akin to a probability distribution. 
Its Fourier transform, the chord function,
or the quantum characteristic function, lies in an alternative phase space.

The reduced Wigner function represents the partial trace 
of a density operator over a subsystem 
and so contains contains all information about a given subsystem.
It results from the projection of the original Wigner function, 
just as in the definition of a marginal probability density.
The reduced chord function is a section of 
the original chord function and also contains all relevant information. 
The purity of the reduced density
operator, i.e. the square of its trace, is a good measure of entanglement,
obtained by integrating either the square of the reduced Wigner function,
or the square-modulus of the reduced chord function.

The Wigner function can be associated to probabilities for measurements
of the eigenvalues, $\pm 1$, of the operators corresponding to 
reflections in phase space, i.e. generalized parity operators. 
These are wavelike properties which appear unintuitive if the
quantum system is pictured as a particle.
Bell inequalities for such measurements can be violated
even for {\it classical looking} states with positive Wigner functions
that have evolved {\it classically} from product states.
These include the original EPR state. 

Entanglement with an unknown environment results in decoherence.
A simple example is that of the centre of mass of a large number
of independent particles, entangled with internal variables.
In this case, the Central Limit Theorem for Wigner functions
leads to some aspects of Markovian evolution for the reduced system.

\end{abstract}

\maketitle

\section{Introduction}

The realization that quantum mechanics admits entangled states
goes back to Schr\"{o}dinger in 1926 \cite{Schroed26}.
He went on to coin the term {\it entanglement} in 1935  \cite{Schroedinger},
but the dramatic example of a biological cat coupled to a decaying nucleus
was never meant to be operational.
Einstein, Rosen and Podolsky (EPR \cite{EPR})
discussed an example of a simple entangled bipartite state in the same year.
Their concern was the compatibility between Heisenberg's indeterminacy principle
and the generation of strong correlations through a measurement on a member of
a pair of particles, even when they could no longer be interacting.
It was the formulation of Bell inequalities, starting in 1964 \cite{Bell} 
that provided a {\it litmus test} for nonlocal correlations
in quantum mechanics. The initial concern was centred on
hidden variable theories and the possibility of their emmulating
quantum correlations even for particles that have ceased to interact.
Such violations of local causality, detected by Bell inequalities, 
could not have developed within any kind of classically evolved ensemble, 
irrespective of whether the variables are explicit or hidden.

Quantum information theory \cite{N-C} has given a new boom to the study
of the qualitative distinctions between classical and quantum mechanics
and to establishing their quantitative measures.
There is nothing so dramatic about the development of nonclassical correlations 
between particles that are still undergoing an interaction, 
but this question has acquired promissing applications in future quantum computations.
Necessarily, these deal with finite dimensional (Hilbert) state spaces,
for which the appropriate entanglement measures are now well established.

One of the difficulties in aplying semiclassical methods
to the study of entanglement is that the former have been developed
for infinite dimensional Hilbert spaces.
Not only are these an extrapolation from the few qubits that have been usually 
considered in quantum information theory,
but entanglement is most clearly exhibited through the correlations in elementary
{\it either-or} experiments. This seems to privledge simple state spaces of a single qubit, 
such as spin-$1/2$ systems. For this reason, Bohm's version of EPR \cite{Bohm}
has become much more popular than the original full phase space version.
A way around this difficulty is to consider the measurement  
of special observables which have only a pair of eigenvalues,
even though they operate on states within an infinite space.
It turns out that one of the most renowned phase space representations in quantum mechanics,
the Wigner-Weyl representation, is based on such operators. Usually, this representation
is viewed as a way of eliciting classical features in a quantum state, 
but it will be used here mainly as a probe into nonclassical correlations.

The development of semiclassical theory throughout the last century
allows us to trace the classical skeleton underlying many features of quantum evolution.
These classical structures are the core of approximations that
improve asymptotically in the limit of large classical actions,
or, more formally, as Planck's constant, $\hbar \rightarrow 0$.
In the case of a finite dimensional Hilbert space, this becomes
the limit of large dimensions.
Even though entanglement is a subtle phenomenon, it leads to gross violation
of inequalities and to quantitive measures which are not beyond the accuracy
of semiclassical approximations. Therefore, it is appropriate to enquire into
the manner in which classical structures can be implicated in such a very
nonclassical feature of quantum mechanics. 

\begin{figure}
\includegraphics[width=12cm]{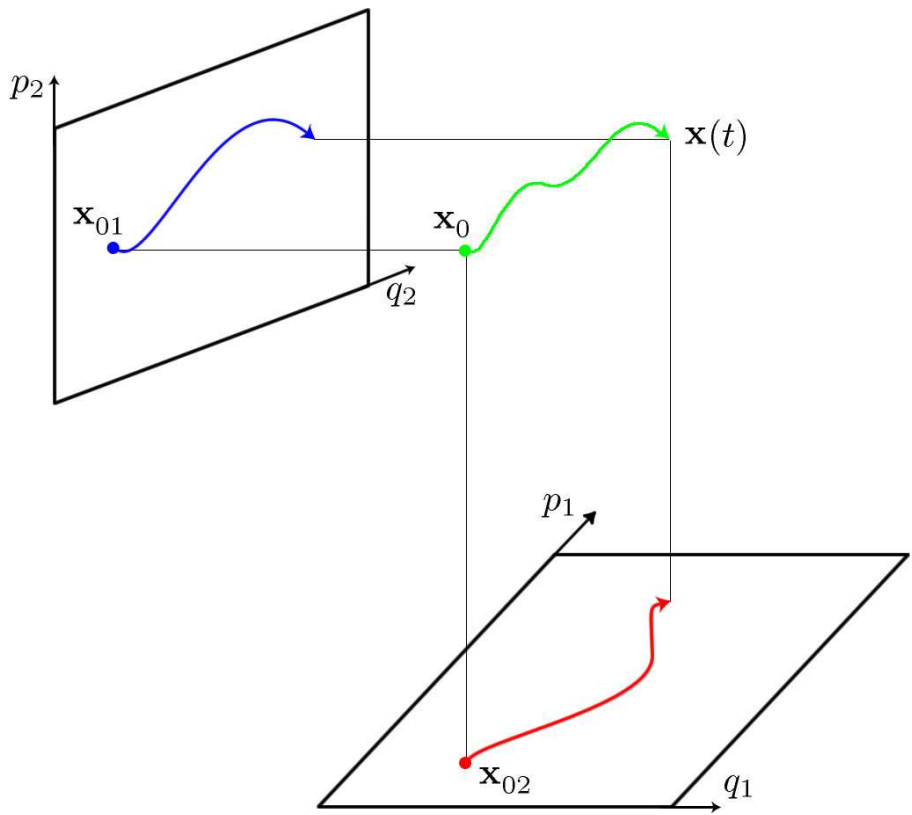}
\caption{The classical trajectories, $\x_1(t)$ and $\x_2(t)$, for each component 
in its own phase space, are projections of the full trajectory for the entire system.}
\label{desenho1}
\end{figure}

Traditionally, semiclassical theory is concerned with the unitary
quantum evolution of closed systems, which are thus described classically
by Hamiltonian dynamical systems. Each point in phase space accounts
completely for the state of the classical system which evolves along a trajectory.
A bipartite, or multipartite system is accomodated in this correspondence
by a higher dimensional phase space. Each point still evolves as a 
single-dimensional (1-D) trajectory, but its projections onto the subspaces, 
which describe the succesion of possible states of each component of the system,
are also 1-D trajectories in their own right, as shown in Fig.~\ref{desenho1}. 

Part of the power of Hamiltonian dynamics
lies in the freedom to transform between different sets of phase space coordinates.
This {\bf canonical invariance} emphasises the importance of the unified evolution
of the full system over that of the component trajectories, 
which are seen to depend on the particular choice of coordinates.
In contrast, it is in the separation into components that 
the phenomenon of quantum entanglement emmerges.  
The particular nature of quantum measurement lies behind
this difference, as is discussed in section 2:
It is only when this is combined to the preceding 
unitary evolution, that the unclassical correlations
between the components become manifest.

Therefore, the study of the properties related to entanglement
should be viewed as an objective that is imposed externally on
semiclassical physics, which perhaps explains the low priority 
received by this goal so far. In these lectures,
we will only be concerned with the most elementary kind of entanglement, 
i. e. that of pure bipartite states, for which the measures of entanglement
are well established. Even so, it will be seen that this simple case
requires the introduction of theoretical instruments of semiclassical theory
that are far from elementary. Not only do we need to cope with a higher dimensional
phase space for the description of a biparite system, 
but it will be shown how the simplest
semiclassical description of operators is achieved in a phase space with
double the dimension of the one corresponding to the states
on which they act. Conversely, some of the most relevant structures
for entanglement, such as partial traces and probability densities, 
can be interpreted as projections of the Wigner function, or sections 
of its Fourier transform. 

The following section reviews the different ways in which 
features of quantum mechanics, interference and entanglement, are
{\it nonclassical}. Simple examples introduce the reflection
symmetries, quite familiar for classical waves, that will play a major role
in the Wigner function formalism. Prior to this though, it is useful
to consider classical-quantum correspondence in a more simple-minded way.
This is the subject of section 3, which introduces product spaces for both
quantum states and classical probability distributions in phase space.  
In either case, the factorizability is broken by an interaction Hamiltonian,
leading to correlations for measurements on the different components.
In the classical case, these correlations are constrained by general Bell inequalities.
This section also introduces the Schmidt decomposition of quantum states. 

Section 4 reviews standard semiclassical theory for quantum states.
Special emphasis is given to products and factorization of both the phase spaces
themselves and the internal {\it Lagrangian} surfaces that support the quantum states.
This product structure is then generalized in section 5 to the representation
of operators. Dyadic operators of posistion or momentum eigenstates form a 
complete linear basis for all quantum operators, 
which correspond to planes in double phase space. 
Linear canonical transformations take these into the
phase space coordinates for the Weyl representation and the chord representation,
its Fourier transform. These bases are associated respectively to
phase space reflections and translations and to the corresponding quantum operators.
In the case of the density operator, we thus obtain the Wigner function
and the chord function, both presented in section 6.
Though the Wigner function cannot be interpreted as a probability
distribution, because it may be negative, it coincides
with the difference for probabilities of measuring either
the positive, or the negative eigenvalue of the reflection operator.
Section 7 is dedicated to projections of the Wigner function and sections
of the chord function, which represent the reduced density operators.
The loss of purity of the latter, obtained as integrals of the square of 
either the reduced Wigner function, or the reduced chord function,
indicate that the overall state is entangled.

It may be guessed that an initially {\it classical} pure state,
the product of Gaussian Wigner functions, would not be entangled 
by a simple rotation of positions and momenta. 
After all, this class of states, including the original EPR states, 
could stand in for a classical phase space distribution.  
However, this is not so, as shown in section 8: 
The reflection correlations for such states violate Bell inequalities, 
even though measurements of positions
and momenta can only correlate classically. 
The transformation to centre of mass coordinates for
any number of particles, studied in section 9, has similar features.  
By invoking the Central Limit Theorem for Wigner functions,
we obtain features of the nonunitary evolution of the centre of mass
in agreement with Markovian theory, i. e. the exact solution of the
Lindblad equation for the density operator.

The final section relates double phase space geometry to
the semiclassical Wigner and chord functions. These are not known
in detail for eigenstates of chaotic Hamiltonians, but it has been proved
that {\it ergodic eigenstates} are  supported by the entire energy shell.
In this case the unitary transformation which factorises the state
can have no classical correspondence.
 
A lot of the experimental work related to entanglement
has been carried out in quantum optics. Rarely is the full generality
of semiclassical states employed there and one can rely mainly on states
derived from the eigenstates of the harmonic oscillator, 
even when phase space is invoked \cite{Schleich}. For this
reason, the initial examples of phase space structures are here
chosen among states of this type. The reader who wishes to avoid
the more subtle aspects of semiclassical theory can mostly skip
sections 4, 10 and parts of 5.

\section{Entanglement and classical physics}

Entanglement is considered to be a quintessential quantum property which
defies all attempts at a classical correspondence. For this reason, 
its description in terms of the classical concept of phase space might
appear foolhardy. It could be that the semiclassical program of
uncovering meaningful relationships between XIX'th and XX'th century
mechanics would be overstretched. Perhaps, though, such an endeavor
would make more sense if it were recalled that the usual validity 
of a classical description of macroscopic phenomena can be attributed 
to the effect of decoherence. In its turn, this results from the entanglement
of a given system with an uncontroled environment, caused by interactions 
that can be minimized, but never entirely eliminated.
Thus, in spite of the fact that the common working languages
employed in classical and quantum mechanics are quite alien to each other,
it is hard to fully comprehend why the outcome of decoherence should be 
the emerging appropriateness of a classical description
for quantum systems, unless we can detect its traces even within
entanglement itself. A simplified version of this program will be
sketched in section 10.

Before attempting to establish a bridge between some features
of quantum entanglement and classical mechanics, it is worthwhile
to consider the more obvious way in which 
interference already separates these theories.
In contrast, the analogy of quantum mechanics with classical waves is much smoother:  
The latter may be superposed linearly and they interfere 
in the same way as matter waves.
In a simple two-slit experiment, the initial quantum state is prepared as a coherent
superposition of momentum eigenstates, 
with eigenvalues that can be {\it classically} measured:
The probability for each momentum direction is the same
as for a uniform ensemble of {\it classical} states.
The evolution through a pair of slits generates classical interference,
equally observable in water waves, or sound waves.
Quantum strangeness only emerges 
if the intensity of the resulting interference pattern
for the conjugate variable, the position, is interpreted 
as the probability for the position measurement of a single particle,
moving according to classical mechanics.
Even so, the particular nature of quantum measurement itself does not
play a prominent role in the phenomenon of quantum interference. The
subsequent quantum state is certainly redefined by the result of the measurement,
but this is not a crucial feature of quantum interference, no matter how unclassical 
its interpretation for a single particle.

The success of the semiclassical treatment of interference phenomena
is no real surprise. If we start from Feynman's path integral formalism \cite{Feynman, Schulman},
quantum evolution is described by a continuum of interfering paths.
Semiclassical theory merely groups these around a few particular classical trajectories
with their Feynman phase. The amplitude of each of these discrete interfering terms is then given 
by a local integration over the continuum of paths.
Classical mechanics takes its part in the theory, 
but there is no limitation to classical phenomena and interference is well described.
Indeed, the role of classical mechanics 
is the same as ray optics in classical wave theory.

In contrast to interference,
the unintuitive nature of entanglement is derived from that of quantum measurement itself.
In no way does this tally with the common sense description of the macroscopic world.
Nothing in our everyday experience prepares us for the collapse of a state that is measured
into one of several possibilities. The common sense presuposition would be 
that the effect of the measurement on the system could and should be made negligible. 
Entanglement highlights this phenomenon in a specially subtle way, because it involves
pairs of measurements on systems with at least two degrees of freedom, or {\bf components}.

If we consider classical waves, or particles,
it would be indeed strange to imagine that such a collapse could result from the
measurement of a subsystem, thus constraining the possible states of the complementary 
subsystem:
It is well known that playing a note
on a piano, i.e. exciting a finite string, will provoke a response on the next octave string.
Here we have two nearly independent systems, stretched strings, weakly coupled by the 
surrounding air. Perhaps, it is better to consider the same note on two nearby pianos, 
so that we consider the interaction of identical systems. The wave form assumed instantaneously
by the pair of strings may be used to describe the state of the whole system, 
or else, we may prefer the Fourier representation, in terms of the eigenstates
for the discrete set of allowed frequencies of each string.
These classical strings are completely analogous to the textbook example
in quantum mechanics of particles moving in 1-D, each in its own box.
But there is no way in which a photograph of one of the piano strings will affect 
the sound produced by the other string, no matter how entangled the quantum analogues
happen to be! Likewise, the measurement of the frequency spectrum of the vibrations
of one of the strings does not oblige it to choose among the various overtones and 
we would be even more surprised if this led to a correlated jump in the other string. 

Yet this is just what we would expect for the analogous quantum system composed
of two particles in their 1-D  boxes, coupled by the same Hamiltonian that may
account for the atmospheric interaction. Such a measurement would single out a discrete energy, 
or equivalently a discrete momentum modulus. Furthermore, in the quantum system, we could 
also measure the position of the particle, with a probability density that is 
specified by the wave intensity. No equivalent interpretation can be
imputed to the classical wave, so that such a {\it position measurement}
would then be devoid of meaning. 
\footnote{It should be remembered that the
classical particle analogy here is not related to the phonons that are generated
by second quantization within each mode.}

Just as there are measurements on a quantum system that are meaningless
for a classical wave, there are others which make no sense for a classical particle.
Consider the excitation of a piano string by the same note, but played on a clarinet.
This has only even harmonics, because it is equivalent to a string that is free
on one side. Then only the even harmonics will be excited in the string,
which will be symmetric about its midpoint. Such an even (or odd) parity, i. e.
the symmetry (or antisymmetry) of the classical stationary wave 
is certainly a measurable property of the analogous quantum state.
Indeed, even a classical wave,
a string that is free on one side, could in principle
be used as a probe to measure directly the even component of the wave, instead of exciting it.
But what would it mean to measure the parity of the corresponding classical particle 
in a box? Generalizations of such {\it parity measurements}, 
distinguishing the eigenvalues of {\it non-mechanical observables}, 
will play a major role in the following discussions of entanglement.

Measurement theory lies outside the scope of a semiclassical treatment.
However, such experimental outcomes will be preceded by (unitary) quantum evolution,
which is not so adverse to a classical description. Indeed the process by which subsystems
become entangled is a {\it preparation} that precedes any quantum measurement. It is only in the
probabilistic interpretation of the subsequent measurement on the system 
that the quantum and the classical viewpoints fundamentally diverge. 
As it happens, standard measures of entanglement require that the components
of the system be completely defined, but do not pre-specify the measurements to be
performed. Thus, the presence of entanglement only indicates the possibility
that some subset of measurements will have nonclassical correlations.
It is precisely this lack of definition with respect to future quantum measurements
that allows space for a semiclassical treatment.   

The study of classical waves displays many of the properties of a simple quantum
system.  Indeed, Rayleigh's \textsl{The Theory of Sound} \cite{Rayleigh} 
anticipates some results later rediscovered
in semiclassical theory. However, each piano string is a system with infinite degrees
of freedom. Though it is not forbidden to consider coupled fields \footnote{Perhaps,
quantum superstring theory will tackle entanglement someday.}, the following
lectures will concern mainly systems with a finite number of degrees of freedom.
In most cases, two degrees of freedom already suffice to discuss the relation between
the concept of entanglement and classical mechanics. So we start with a review of
classical-quantum correspondence.

\section{Classical-quantum correspondence}

The simplest quantum systems with a classical correspondence have a single degree of freedom,
e.g. a particle constrained to move in a straight line. The classical state
of the system is described by its position, $q$, and its momentum, $p$. 
Together  they define a point in phase space, $x=(p,q)$, which is a 2-D plane.
Perhaps, classical state space would be a more appropriate term, 
because each point specifies all future motion of a classical system, 
once the Hamiltonian, $H(x)$, is specified, through Hamilton's equations:
\begin{equation}
\dot{p} = -\frac{\partial H}{\partial q} \quad , \quad \dot{q} =
\frac{\partial  H}{\partial p}\ .
\end{equation}
These equations may be compactified into the form
\begin{equation}
\dot{x} = {\J} \frac{\partial  H}{\partial x}\ ,
\end{equation}
with the definition of the $(2\times 2)$-dimensional matrix 
\begin{equation}
{\J} =\left[
\begin{array}{c|c}
0 & -1 
\\ \hline
1 & 0
\end{array}
\right],
\label{J}
\end{equation}
acting on the phase space points, $x=(p,q)$.
Unless $H(x)$ is quadratic, this motion is nonlinear.

Corresponding to this 2-D plane, quantum mechanics matches 
the states $|\psi\rangle$
of an infinite dimensional Hilbert space, ${\bi H}$,
on which act the operators, $\widehat q$ and $\widehat p$. 
Each eigenstate of $\widehat q$, labled by the eigenvalue $q_0$, 
corresponds to the vertical line, $q=q_0$,
whereas the horizontal phase space lines are matched by eigenstates of $\widehat p$.
These operators do not commute,
$[\widehat p, \widehat q]=i \hbar$, but if we appropriately symmetrize the
order in which $p$ and $q$ appear in $H(x)$, then the motion of the states
$|\psi\rangle$ is also determined by the quantum Hamiltonian $H(\widehat x)$, 
through the linear equation,
\begin{equation}
i \hbar \frac{\partial}{\partial t}|\psi\rangle = H(\widehat{x})|\psi\rangle ,
\end{equation}
i.e. Schr\"{o}dinger's equation.

The uncertainty principle excludes the existence of a quantum state
that corresponds precisely to a phase space point.
However, the unavoidable dispersion in measurements of position,
or momentum allow us to seek an approximate
correspondence with probability distributions of phase space points.
This is unsatisfactory as far as interpretation is concerned, 
because probabilities are matched to the square of a state rather than the state itself.
Nonetheless, a certain intuition can be obtained through this analogy.
Given a phase space probability density, $f(x)$, the expectation value
of any {\bf classical observable} $O(x)$ is given by
\be
E(O)=\int {\rm d}x\>O(x)\>f(x).
\label{clexpectation}
\ee
Hence, the dispersions in position and momentum are 
$\delta q^2=E((q-E(q))^2)$ and $\delta p^2=E((p-E(p))^2)$.
The uncertainty principle then imposes that only phase space distributions
for which $\Delta'=\delta q \delta p \geq \hbar$ should be considered.

However, this quantity is not a classical invariant. 
The flow, $x(0) \rightarrow x(t)$, generated by the Hamiltonian is 
a {\bf canonical transformation}, so that \cite{Arnold}, for all $t$,
\be
\oint_{\gamma_0} p(0) \cdot \rmd q(0) = \oint_{\gamma_t} p(t) \cdot \rmd q(t),
\label{canonical}
\ee  
where $\gamma_0$ is any circuit and $\gamma_0\rightarrow\gamma_t$.
General Hamiltonian evolution will stretch and bend any closed curve that
is initially compact, so that a probability distribution that is unity inside
$\gamma_0$ and zero outside will not have constant $\Delta'$. 
Linear canonical transformations, that is, {\bf symplectic transformations},
are well known to be specially favourable for classical-quantum correspondence,
as will be further discussed. It will be shown in section 6 that, $\Delta$,
the determinant of the {\bf covariance matrix},
\begin{equation}
{\bf K} =\left[
\begin{array}{c|c}
\delta p & \delta pq
\\ \hline
\delta pq & \delta q
\end{array}
\right],
\label{Kov}
\end{equation}
where $(\delta pq)^2=E(pq-E(p)E(q))$, is invariant under symplectic transformations.
 
To discuss entanglement, we need more than one degree of freedom. Quantum 
states can then be decomposed into a basis of product states, 
\be
|\psi\rangle=|\psi_1\rangle \otimes...|\psi_l\rangle \otimes...|\psi_L\rangle,
\ee
which span the full Hilbert space, ${\bf H}={\bi H}_1\otimes...{\bi H}_l\otimes...{\bi H}_L$,
i.e. the tensor product of the factor Hilbert spaces that describe each degree of freedom.
Likewise, the full phase space is now a Cartesian product of 2-D conjugate planes,
each the phase space for a particular degree of freedom,
\be
x=x_1 \times...x_l \times...x_L
\ee
and thus has $2L$ dimensions. However, we must be wary of the difference between
the classical and quantum geometries: a phase space strip, $\delta q$, corresponds
to this range of eigenvalues for the operator $\widehat q$. 
This set of eigenstates spans an infinite dimensional subspace of the product Hilbert space, 
whatever the number of degrees of freedom. On the other hand, each of these
position eigenstates corresponds to one of the parallel $L$-D $q$-planes  
within the $2L$-D phase space strip. 

The classical or quantum motion for systems with more than one degree of freedom is still
defined by a Hamiltonian, $H(x)$, or $H(\widehat x)$, but now ${\partial  H}/{\partial x}$
is a $2L$-dimensional vector and $\J$ is a block matrix. We shall also use
the {\bf skew product},
\be
x\wedge x'=\sum_{n=1}^L (p_l q'_l - q_l p'_l)= \J\>x \cdot x'. 
\label{squew}
\ee
This {\bf symplectic area} of the parallelogram formed by the vectors
$x$ and $x'$ is invariant with respect to symplectic transformations.
Again, these are linear canonical transformations, with (\ref{canonical})
interpreted as a line integral in the $2L$-D phase space. For higher dimensional systems,
all even dimensional volumes, from 2 to $2L$, 
are preserved by canonical transformations \cite{Arnold}.

 %
%

If the degrees of freedom are completely decoupled, each with its own 
probability distribution, $f_l(x_l)$, the full probability distribution
will be just the product,
\be
f(x)=f_1(x_1)...f_l(x_l)...f_L(x_L).
\ee
In this case, the probability distribution for a single degree of freedom
is reobtained by {\it tracing over} the other variables:
\be
f_1(x_1)=\int f(x)\>\rmd x_2...\rmd x_L.
\label{marginal}
\ee
If the full probability distribution cannot be factored into 
a product, the above equation then defines the {\bf marginal distribution}.
This process foreshadows that of partial tracing over the density operator, 
to be studied in section 7, which is central to the study of entanglement.

The product nature of the classical distribution will be retained
throughout the evolution, if the full classical Hamiltonian is purely additive,
\be
H(x)=H_1(x_1)+...H_l(x_l)+...H_L(x_L),
\ee
i.e. if there is no coupling between the motions of the several degrees of freedom.
This follows from the decoupling of Hamilton's equations into
\be
\dot x_l=\J_l \frac{\partial H_l}{\partial x_l}, 
\ee
for each degree of freedom.
In other words, if $k \neq l$, then $x_l(t;x_{l0})$ does not depend on $x_k$
(nor on the initial value, $x_{k0}$).
Furthermore, we then have
\be
f_l(t;x_l)=f_l(t;x_l(-t;x_l))\>\> {\rm and} \>\>f(t;x)= f_1(t;x_1)...f_L(t;x_L),
\ee
where $x_l(-t;x_l)$ specifies the past location of $x_l$.
Likewise, the volumes in each subspace will be preserved
and the conservation of the $2L$ dimensinal volume is just that
resulting from the conservation of the factor volumes.

For a classical system, the transition from product probabilities
to general probabilities can only be generated by coupling
terms in the driving Hamiltonian, containing cross products, 
which are at least bilinear in the different variables.
A general classical observable will be a function of all the phase space
variables and its expectation is accordingly given by (\ref{clexpectation}).
For instance, this might be the {\bf either-or observable}, $O_1=\pm 1$ for detecting some 
physical property associated with one particle, or the detection of $O_2=\pm 1$
for a second particle. For classical particles which have been allowed
to drift sufficiently far from each other after interacting, 
the result of the $O_1$-measurement
will not affect the $O_2$-measurement and vice versa. Therefore,
the correlation must be represented in the form
\be
E(O_1;O_2)= E(O_1\>O_2)=\int {\rm d}x_1{\rm d}x_2\>O_1(x_1)\>O_2(x_2)\>f(x_1, x_2).
\label{local}
\ee  

This equation has the same form as correlations postulated for
{\it local hidden variable theories} \cite{Bell, Peres}. 
Indeed, one of the reasons for this choice
is that (\ref {local}) must hold for any evolution of $f(x)$ governed by classical mechanics. 
This form for the correlation between different components of the system 
is then taken as a prerequisite
for theories that in all other respects should give the same results as quantum mechanics.
Since this is certainly not one of the objectives of classical mechanics,
such conjectures then necessarily demand extra, unknown and hence {\it hidden} variables.

It is due to the seminal work of Bell \cite{Bell} that we are able
to compare, through inequalities, the correlations predicted by quantum mechanics 
with a very wide range of possible local correlations. 
The point is that any measurement
affects the entire quantum state, i. e. both its components, 
unless the state happens to be an eigenstate of the measured observable.
So quantum measurements are not local in the sense that led to (\ref{local}).
In case of the general CHSH inequality \cite{CHSH, N-C},
involving either-or observables, $O_{1a}$, $O_{1b}$, $O_{2a}$ and $O_{2b}$,
(\ref{local}) implies that
\be
|E(O_{1a};O_{2a})+E(O_{1a};O_{2b})+E(O_{1b};O_{2a})-E(O_{1b};O_{2b})| \leq 2.
\label{CHSH}
\ee 
As well as constraining possible hidden variable theories,
this inequality can be used as a detector of nonclassical correlations
in quantum mechanics. This kind of nonclassicality, {\bf entanglement},
is much more subtle than quantum interference effects, as will be discussed
in the later sections.
A dip into \textsl{Bertlmann's socks and the nature of reality} \cite{socks}
by Bell provides a delightful discussion of all the main points concerning
classical locality versus quantum correlations.
The book by Peres \cite{Peres} is also recommended.

It is worthwhile to discuss some specific examples of systems with more than
one degree of freedom. An obvious possibility is a collection of particles,
each moving in one dimension. Another is a single particle moving in two,
or three dimensions. Classical and quantum mechanics make no distinction
between these alternative interpretations of the dynamical variables.
All that is demanded is that the variables pertaining to different 
degrees of freedom commute, $[\widehat p_k,\widehat q_j]=i\hbar\delta_{kj}$,
or, correspondingly, that the classical Poisson bracket 
$\{p_k,q_j\}=\delta_{kj}$ (see e.g. \cite{Goldstein}).
We can also use angular momentum and
their conjugate angles. But are other variables, 
obtained through classical canonical transformations, allowed?

For example, consider our piano string, now modeled as $L$ masses connected
by harmonic springs. We can switch to the $L$ normal modes of vibration.
This is a linear canonical transformation, which substitutes the original
$L$ conjugate planes, $x_l=(p_l,q_l)$, by new conjugate planes, $x'_l=(p'_l,q'_l)$,
that now describe collective motions of the $L$ masses. This is also a
proper phase space to be quantized, $x'_l\rightarrow\widehat x'_l$.
Another important example of a quantizable canonical transformation
follows from the description of a collection of particles in terms of the
centre of mass together with internal coordinates.

Whatever the physical realization, symplectic transformations, 
correspond exactly to unitary quantum transformations and hence to equivalent
quantum systems \cite{Vor76}. These transformations generally redefine the components
of the full system and may take an entangled state into a product state,
or vice versa. Any measure of entanglement is affected 
by such a general transformation, so one requires only that the measure be invariant
with respect to local unitary transformations, lying within each separate component.  
As for nonlinear canonical transformations, these are not exactly
matched by quantum unitary transformations \cite{Vor76} and, 
hence, cannot be directly quantized.
It might still be useful sometimes to push this correspondence through, but it must be
remembered that the result is only a semiclassical approximation.

Taking again the continuum limit, $L\rightarrow\infty$, each normal mode of the
finite chain converges onto one of the lower modes of the continuous string.
There is no essential difference between the interaction
and hence the entanglement among these modes of the continuum 
and that of finite modes (caused by residual nonquadratic terms in the Hamiltonian). 
In each case, there corresponds a plane in the phase space, which is of infinite dimension
in the case of a field. The entanglement between modes of the electromagnetic field
within a finite cavity also has a similar interpretation in terms of a classical field.
The unperturbed motion is now that of a quantized harmonic oscillator, 
corresponding to a classical oscilation in each phase plane . 
 
Another example is that of a particle with {\it internal structure}.
The latter may be described by an angular momentum, coupled
to the translational degrees of freedom by an external field.
The Stern-Gerlach experiment is just such a system, in which the 
magnetic moment, tied to the spin angular momentum of the electron
is coupled to its position by an inhomogeneous magnetic field.
The spin is an intrinsically quantum mechanical two level system
and the interest in quantum information theory tends to emphasise
such simple quantum systems. But, in principle, there is no difference
between this case and a Rydberg atom, prepared in a state with a large electric 
dipole moment, coupled to position through an inhomogeneous electric field.
Such a system can be described more naturally in classical terms.
Cavity quantum optics deals with the coupling and hence the entanglement 
of the internal states of individual Rydbreg atoms with a specific mode of the
electromagnetic field.

For all these systems, coupling terms in the overall Hamiltonian
will destroy the product form of an initially decoupled quantum state,
or classical distribution. We should bear in mind three basic differences between
classical and quantum systems: i) the nature of the initial state;
ii) the nature of the evolution and iii) the effect of experiments.
As we have seen, the last is the most radical difference, which, indeed,
gives rise to the concept of entanglement. Our objective here is to cast the
quantum mechanical description of i) and ii) in the most classical terms possible,
so as to highlight the truly inovitative elements of the quantum theory
when iii) is considered.

A fundamental difference between the quantum and classical descriptions
should be discussed before proceeding: The analogy between the evolution
of classical probability distributions and quantum states is somewhat deceptive
in as much as the latter determine only probability amplitudes which can be
complex and interfere with each other. To arrive at a closer analog of
probabilities, we should, in some sense, square the quantum states. 
The correct procedure is to define density operators, 
or their phase space representation, Wigner functions,
to be studied in section 6.
However, their evolution is nonclassical, unless the Hamiltonian is quadratic.

It will be only in the context of the density operator that it becomes
meaningful to distinguish between pure states and mixed states.
Taking an average over a set of probability distributions
defines a new probability distribution. 
Likewise, if we superpose the corresponding quantum states, $|\psi_j\rangle$,
we obtain a new quantum pure state. But if we average over the coresponding
{\bf pure state  density operators}, $\rho_j=|\psi_j\rangle\langle\psi_j|$, 
there results a mixed state. The latter will be discussed in section 6. 

Now it is important to bring out
a special form of state superposition. This is the {\bf Schmidt decomposition},
\be
|\psi\rangle=\sum_j \lambda_j \> |\psi_1\rangle_j \otimes |\psi_2\rangle_j,
\label{Schmidt}
\ee
which exists for any bipartite state (see e.g. \cite{N-C}). It must be emphasised that both factor states
in the above tensor products may themselves correspond to several degrees of freedom,
but the result is only proved if there are only two of them.
The product states form a particular orthonormal basis in which to describe
the state, $|\psi\rangle$, so that the real, non-negative coefficients,
$\lambda_j$, satisfy $\sum_j {\lambda_j}^2 =1$. 
The state is entangled, unless $\lambda_j=\delta_{1,j}$.
The Schmidt decomposition is often employed for the description of entangled states
in finite Hilbert spaces. In this case, the number of nonzero eigenvlues, $\lambda_j$,
is a relevant quantifier of entanglement, known as the {\bf Schmidt number}.
For infinite dimensional Hilbert spaces, there may be an infinite 
number of nonzero Schmidt coefficients.

\section{Semiclassical quantum states}

Consider a momentum eigenstate $|p'\rangle$ for $L=1$.
In the momentum representation, this is just
\be
\langle p|p'\rangle = \delta (p'-p),
\ee
which is not in a good form for semiclassical extrapolation. 
For this purpose it is better to use the complementary representation,
\be
\langle q|p'\rangle = \exp\left(\frac{iqp'}{\hbar}\right)=\exp\left(i\frac{S_{p'}(q)}{\hbar}\right).
\ee
The phase in this expression can be interpreted as the area between the classical curve 
(the straight line, $p'=p$) and the $q$-axis. There is also an arbitrary constant phase, 
which is established by the choice of the initial point for the integral,
\be
S(q)=\int_{q_0}^q p(q)\> \rm d q.
\label{action}
\ee

\begin{figure}
\includegraphics[width=11cm]{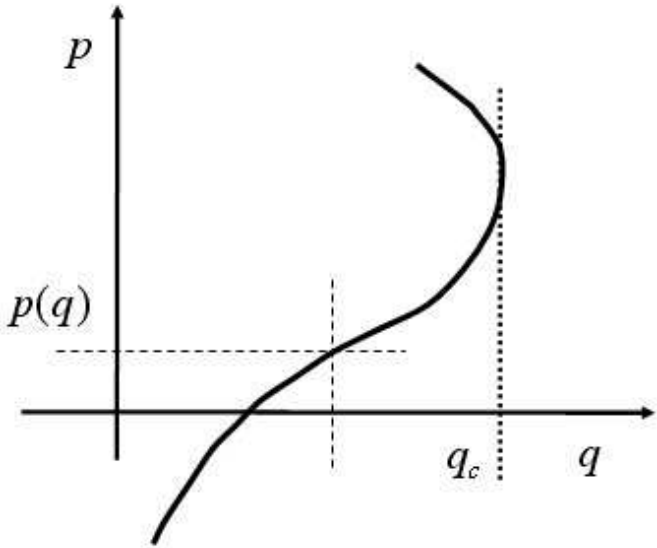}
\caption{The caustic of the semiclassical approximation to 
$\langle q|k\rangle$ lies in the neighbourhood of the point $q_c$,
where the tangent to the classical curve is vertical. 
The projection of this region onto the p axis is nonsingular,
leading to a good semiclassical approximation for $\langle p|k\rangle.$}
\label{caustic}
\end{figure}

Consider now a general observable, $K(\widehat p,\widehat q)$. 
Its eigenstates correspond classically to curves, $\gamma$, in phase space: 
$K(p,q)=k$. These may be viewed locally as (possibly multivalued)
functions, $p_j(q)$. Then the simplest semiclassical approximation is
\be
\langle q|k\rangle = \sum_j A_j(q) \exp\> i\left[{S_j(q)\over \hbar}+\nu_j\right],
\ee
see e.g.\cite{livro}.
The phases, $S_j(q)$ are again obtained from (\ref{action}).
The extra constant phases, $\nu_j$, are known as {\it Maslov indices} 
\cite{Gutzwiller, livro}, but they will not be discussed here. 
The amplitudes, $A_j(q)$, are defined purely in terms of the
classical structure. They are finite 
wherever the vertical line, $q= constant$, intersects the classical curve
transversely. Where this vertical line is tangent to the classical curve, 
such as $q_c$ in Fig.\ref{caustic}, the amplitude diverges.
These points where the semiclassical approximation breaks down are known
as {\bf caustics}. The different branches of the function $p_j(q)$ are connected
at caustic points. 

In the case of bound eigenstates of $\widehat K$,
the curves, $\gamma$, are closed. Then the eigenvalues 
are approximately obtained by the Bohr-Sommerfeld quantization condition,
\be
\oint_{\gamma}p\> \rmd q=(n+{1\over 2})\hbar.
\ee
The quality of the semiclassical approximation for both the states themselves
and their eigenvalues improves for large quantum numbers, $n$.
Ground states, including that of the harmonic oscillator, 
are badly described by these approximations.  

Even for large $n$, a closed curve, $\gamma$, must inevitably
have at least a pair of caustics. The way around this is to switch
to the $p$-representation. Then the vertical tangent at the caustic
position, $q_c$,  shown in Fig.\ref{caustic}, 
would correspond to the state, $\langle p| q_c\rangle$,
which is in a nice semiclassical form. This means that the local branch
of the multivalued function, $q(p)$, gives rise to a semiclassical approximation
which is a superposition of terms of the form
\be
\langle p|k\rangle = B(p) \exp\> i\left[{S(p)\over \hbar}+\nu\right].
\ee 
This allows us to define the correct semiclassical approximation in the
$q$-representation through the caustic region by the Fourier transform
\be
\langle q|k\rangle ={1\over(2\pi\hbar)^{1/2}} 
\int \rmd p\> \langle p|k\rangle \exp\left(\frac{iqp}{\hbar}\right),
\ee
which leads to a more refined approximation in terms of Airy functions 
instead of exponentials. This is usually refered to as the Maslov method 
of dealing with caustics \cite{Maslov} (also discussed in \cite{livro}). 

Let us now consider a product state for $L>1$. Then,
\be
\langle q|p'\rangle = \exp\left(\frac{iq_1p'_1}{\hbar}\right)...\exp\left(\frac{iq_Lp'_L}{\hbar}\right)
=\exp\left(\frac{iq\cdot p'}{\hbar}\right)
\ee
and we can generalize the definition of action,
\be
S(q)=\int_{q_0}^q p(q)\cdot \rmd q.
\label{actionL}
\ee
This does not depend on the choice of path between $q_0$ and $q$,
because $p'(q)$ is a constant in this simple case. Hence,
this function defines a {\bf Lagrangian surface}, i.e. a surface
such that
\be
\oint p\cdot \rmd q=0,
\ee
for any (reducible) circuit \cite{Arnold}.

In general, the product state will involve arbitrary eigenstates
of $L$ observables, $\widehat K=\widehat K_1\widehat K_2...\widehat K_L$,
each in its own Hilbert space:
\be
\langle q|k\rangle =\langle q_1|k_1\rangle ...\langle q_L|k_L\rangle. 
\ee
The wave function will be a superposition of terms with the form
\be
\langle q|k\rangle = \prod_l A_l(q_l) \exp\> {i\over \hbar}\left[S_1(q_1)+...+S_L(q_L)\right],
\label{scproduct}
\ee
one term for each branch of the funtions, $p_l(q_l)$.
 
Defining again $S(q)$ as the above phase, it is seen to be independent of the order in which 
we progress along each segment $(q_{0l},q_l)$, 
while keeping the other integration variables constant:
The definition (\ref{action}), now reinterpreted as a path integral, 
is independent of the path on the surface.
Therefore, this more general surface, $K(p,q)=k$, is also Lagrangian.

If the surface is the product of $L$ quantized {\it circles} (closed curves),
it will be an {\bf $L$-torus}, $\tau$. Each of the $L$ irreducible circuits, 
$\gamma_l$, must then satisfy the Bohr-Sommerfeld conditions,
\be
\oint_{\gamma_l}p_l\cdot \rmd q_l=(n_l+{1\over 2})\hbar,
\ee
or some suitable generalization (see e. g.  \cite{livro}). 
Notice that the line-integral here
used is not restricted to plane sections of $\tau$, because all
topologically equivalent circuits on a Lagrangian surface must have the
same action.

Let us now evolve the product state semiclassically. The basic result,
due to van Vleck \cite{Van Vleck}, can be reinterpreted as the statement that
classical and quantum evolutions {\it commute}. In other words, 
we can evolve classically each curve, $\gamma_l$, 
if there are no cross terms in the Hamiltonian,
so that the different degrees of freedom are decoupled.
Each evolved observable then corresponds to $K_l(x_l,t)=K_l(x_l(x_{l0},t),0)$
and we approximately reconstruct the classically evolved state from the evolved torus, 
$\tau_l(t)$, which is the product of the $\gamma_l(t):K_l(x_l,t)=k_l$. 

Notice that this 
classical evolution of products of curves fits in to the general view
concerning the evolution of product probability distributions in the
previous section, by merely choosing $f_l(t;x_l)=\delta(K_l(x_l,t)=k_l)$
and running time backwards.
The important distinction between classical and semiclassical evolution is that
the latter contains interferences between the different branches of the evolving
classical curve. Each representation exhibits these interferences in a different way.

Just as cross terms containing products of the different variables in the 
Hamiltonian destroy the product form of a classical probability distribution,
the classically evolved $L$-D surface corresponding to an original product state
also ceases to be a product. However, the smoothness of the evolution implies that
the topology of the surface must be preserved(be it plane, torus, or, in between: cylindrical) .
Furthermore, the classical evolution, $x_0\rightarrow x_t$,
is a canonical transformation, and hence 
all reducible circuits on the evolved surface have zero action, 
i.e. $\tau_t$ still has the
Lagrangian property, which allows us to define the path-independent action $S(q)$,
and the irreducible circuits of $\tau_t$ 
still satisfy the same Bohr-Sommerfeld conditions, to first order in $\hbar$.

\begin{figure}
\centering
\includegraphics[width=11cm]{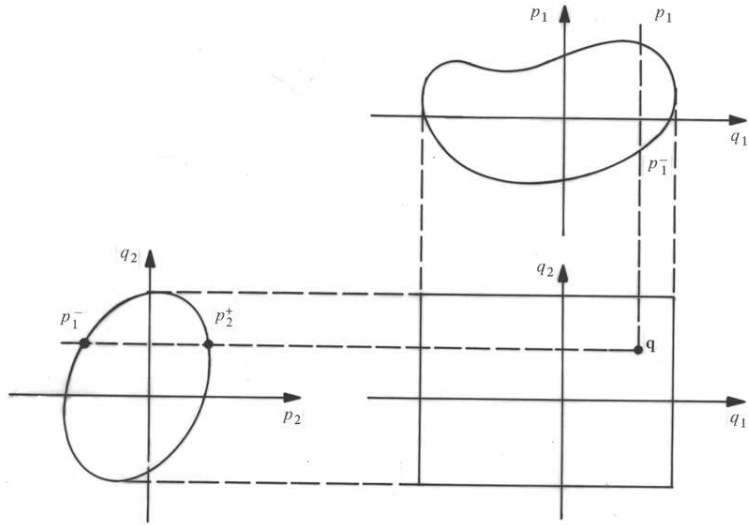}
\caption{Each point within the rectangular caustic of a two-dimensional product torus
is the immage of four phase space points under p-projection.}
\label{rectangle}
\end{figure}

Let us investigate further the case of two degrees of freedom.
The separable torus, $\tau=\gamma_1\otimes\gamma_2$, 
can be pictured through the separate $\gamma_1$
and $\gamma_2$ curves. These coincide with sections of the 2-D
torus by alternative 3-D planes.
(The normal case for {\it Poincare sections}, see e.g.\cite{livro})
The $\gamma_2$ curve does not depend on the choice of the $q_1=constant$ section.
The separable torus projects as a rectangle onto position space $(q_1,q_2)$,
as shown in Fig.\ref{rectangle}. Within this rectangle, there are four different branches 
of the torus, which project onto each position, $q$, corresponding to the
combinations of the two branches of each circle. The caustics at the side
of the rectangle are {\it double fold lines}.

\begin{figure}
\centering
\includegraphics[width=11cm]{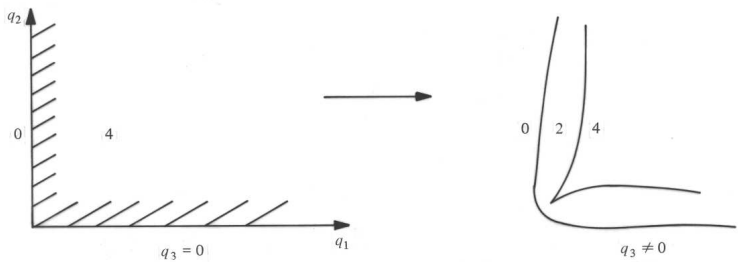}
\caption{Catastrophe theory establishes the generic form for 
the unfolding of the double fold caustic at each corner of the projection
of a product torus as it evolves.}
\label{corner}
\end{figure}

\begin{figure}
\centering
\includegraphics[width=11cm]{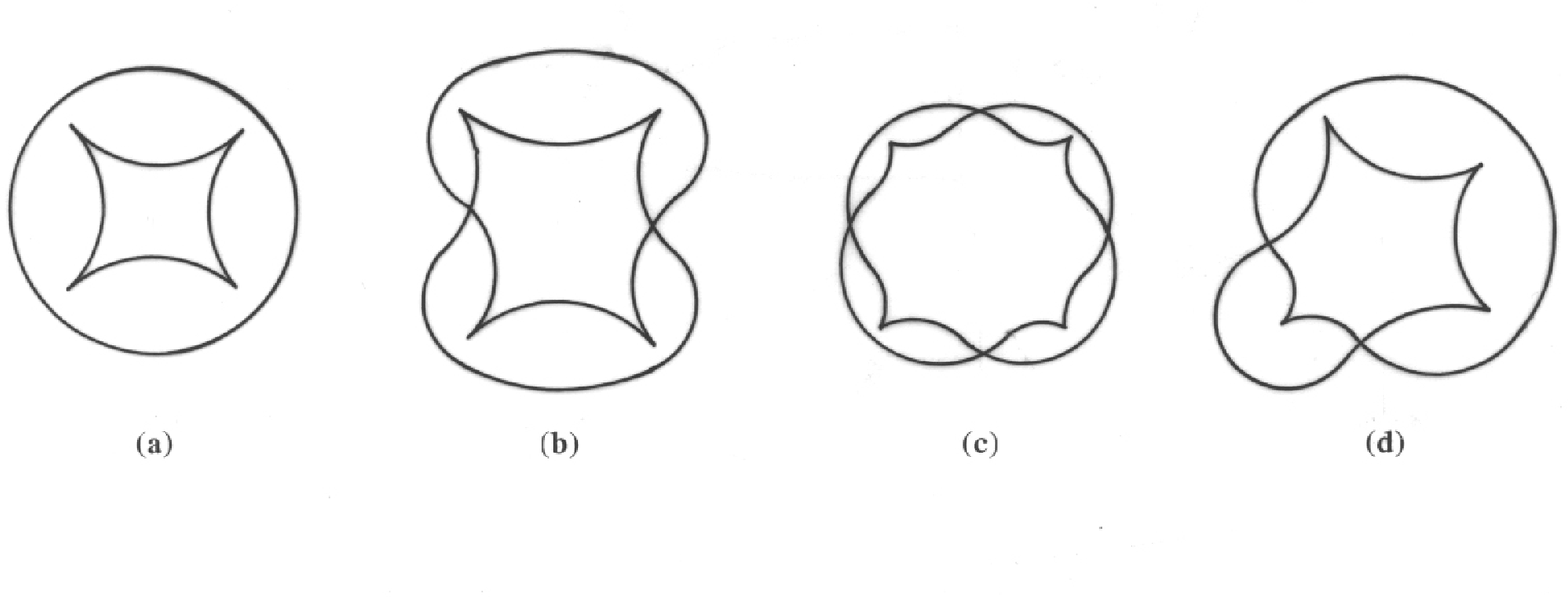}
\caption{The full topology of the full fold lines is not determined
by catastrophe theory: Each of the above forms corresponds to a
different symplectic evolution from an initial product torus. }
\label{unfold}
\end{figure}

After a general canonical evolution, the sections of $\tau$ are no longer equal
for different  choices of $q_1=constant$ (or $q_2=constant$), though all the sections
have the same area, $S_1$ (or $S_2$). In some cases (to do with time-invariance of the Hamiltonian)
the projection onto the $q$-plane will merely distort the rectangle, which
will still have finite-angled corners connecting double fold lines. 
But in general, these corners, {\it hyperbolic umbilic points}, will unfold in the generic
form specified by catastrophe theory, as shown in Fig.\ref{corner}.
There are four possibilities for the topology of the unfolding of the
rectangle, shown in Fig.\ref{unfold}. 
For $L>2$, the projection of the $L$-torus onto the $L$-D $q$-plane
will be a solid hypercube that will be distorted, or unfolded by the motion
generated by a coupling Hamiltonian. (These geometries are reviewed in \cite{livro},
but are more thoroughly discussed in \cite{AlmHan82}.)


The representations of quantum states in terms of orthogonal position, 
or, alternatively, momentum eigenstates are the best that we can do,
because of Heisenberg's uncertainty principle. Semiclassically, this
corresponds to viewing a Lagrangian surface through a set of Lagrangian planes
that {\it foliate} phase space. We switch from the $q$-representation to the 
$p$-representation by means of a Fourier transform of $\langle q|\psi\rangle$.
This corresponds classically to taking the {\it Legendre transform} of $S(q)$ \cite{Arnold}.
For $L>1$, we may take the Fourier transform 
for a subset of the degrees of freedom.
This corrresponds to using a classical description in terms of the
alternative Lagrangian planes $(p_1,...,p_l,q_{l+1},...,q_l)$.

One way to achieve a full phase space description is to use the basis
of {\bf coherent states} \cite{Glauber, KlauderSka, Perelomov, Schleich, Tannoudji}, 
labeled by the phase space vector, $\eta=(\eta_p,\eta_q)$,
\begin{eqnarray}
\langle q|\eta\rangle= \Big(\frac{\omega}{\pi \hbar}\Big)^{1/4}\exp\Big(-\frac{\omega}
{2\hbar}(q-\eta_q)^2+i\frac{\eta_p}{\hbar}(q-\frac{\eta_q}{2})\Big).
\end{eqnarray}
Even though the coherent state basis is overcomplete, the exact decomposition, 
\begin{eqnarray}
|\psi\rangle=\frac{1}{\pi}\int \rm d\eta |\eta\rangle \langle\eta|\psi\rangle,
\end{eqnarray}
is unique.
The coherent states are phase space translations of the ground state of
the harmonic oscillator (with unit mass):
\begin{eqnarray}
\langle q|0\rangle= \Big(\frac{\omega}{\pi \hbar}\Big)^{1/4}\exp\Big(-\frac{\omega}
{2\hbar}q^2\Big).
\end{eqnarray}
These result from the action of the {\bf translation operator}:
\begin{equation}
\label{transop}
\hat{T}_{\eta} =   \exp \big[ \frac{i}{\hbar}
                \left( \eta_p \cdot \hat{q} -
                       \eta_q \cdot \hat{p}
               \right)\big]\ =
\exp {\left({i\over \hbar} \eta \wedge \widehat x\right)},
\end{equation}
using the skew product (\ref{squew}).
If either $\eta_p=0$, or $\eta_q=0$, we obtain the usual
translation operators for momenta, or positions, respectively.
The arbitrary phase due to noncommutation of $\widehat p$ and $\widehat q$
is here chosen in the most symmetric way, using the Baker-Hausdorf relation \cite{Messiah}.

In quantum optics it is customary to switch to the basis of creation and
anihilation operators 
$(\hat{q}\pm i\hat{p})/\sqrt{2 \hbar}$. 
In this context, the translation operator (\ref{transop}) 
depends on the 
complex chords $(\xi_p \pm i \xi_q)/\sqrt{2 \hbar}$ and is called the 
{\it displacement operator} \cite{Glauber}.
The semiclassical limit for a complex phase space is not as transparent 
as the real theory treated here. 
However, it is quite feasible to effect phase space translations in 
an experimental optical context. \cite{LutterbachDav}.

The coherent state representation is not orthogonal and is overcomplete.
The alternative, to be explored in the next section, is to work directly
with operators: We represent operators in {\it orthogonal} operator bases
in analogy to the way that quantum states are commonly decomposed.
This allows us to work directly with the translation operators,
without having to apply them to the ground state of the harmonic
oscillator.

\section{Operator representations and double phase space}

The linear operators, $\widehat A$, that act on the quantum Hilbert space
form a vector space of their own: $|A\rangle\!\rangle$. Defining the {\bf Hilbert-Schmidt product},
\be
\langle\!\langle A|B\rangle\!\rangle= {\rm tr}\> \opA^{\dagger}\widehat B,
\label{HJprod}
\ee
we find that the dyadic operators $|Q\rangle\!\rangle=|q_-\rangle \langle q_+|$
form a complete basis, i.e.
\be
\langle\!\langle Q|A\rangle\!\rangle=\langle q_+|\opA|q_-\rangle=
{\rm tr}\>|q_-\rangle\langle q_+|\opA,
\ee
provides a complete representation of the operator $\widehat A$.
Here, $\opA^{\dagger}$ is the adjoint of $\opA$. 
One should note the similarity between this dyadic basis, $|q_-\rangle \langle q_+|$,
in the case of $L=1$ with the basis of product states, $|q_1\rangle \otimes |q_2\rangle$.
The substitution of a bra by a ket in the former, 
will in most cases imply no more than complex conjugation.

Thus we may relate the vector space of quantum operators 
to a {\it double Hilbert space} with respect to that of quantum states.
Since we have explored the correspondence of the state-Hilbert space
with classical phase space, it is now natural to relate the double Hilbert space
to a double phase space : 
$X=x_- \times x_+$ (se e.g.\cite{Littlejohn95}). The operator $|Q\rangle\!\rangle$
should then correspond to the Lagrangian plane, $Q=constant$ in the double phase space.
This does hold, within a minor adaptation, analogous to the use of the adjoint operator 
in the definition of the Hilbert-Schmidt product. That is, we should define
$Q=(q_-, q_+)$, but $P=(-p_-, p_+)$ as coordinates of the double phase space $X=(P,Q)$.
 
A good reason for this is that then we include among the set of Lagrangian surfaces in
double phase space all the canonical transformations in single phase space,
$x_- \rightarrow x_+=\C(x_-)$. This also transports closed curves, 
$\gamma_- \rightarrow \gamma_+$, so that
we may rewrite the definition of a canonical transformation \eref{canonical} as
\be
\oint_{\Gamma} P\cdot \rmd Q = 0,
\ee
where $\Gamma=(\gamma_-,\gamma_+)$. Thus we may consider $\gamma_\pm$ as projections of 
the curve $\Gamma$ defined on the $(2L)$-dimensional surface,
$\Lambda_\C$, which specifies the canonical transformation, 
within the $(4L)$-dimensional double phase space, $X=(P,Q)$.

It is worthwhile to consider the richness of structures in double phase space. 
On the  one hand, a canonical transformation defines a Lagrangian surface as
$x_+(x_-)$, a one-to-one function. On the other hand, the product of a Lagrangian surface,
$\lambda_-$ in $x_-$ with another surface $\lambda_+$ in $x_+$, 
$\Lambda=\lambda_- \otimes \lambda_+$, is also Lagrangian in double phase space,
but projects singularly onto either of the factor spaces.
In the case that both sufaces are tori, we obtain a double phase space torus,
$\tau =\tau_- \otimes \tau_+$, as if we had doubled
the number of degrees of freedom. 
(All Lagrangian surfaces will hereon be labled $\tau$, 
even when they are not necessarily a torus; in the case that $L=1$,
$\tau$ is just a closed curve, $\gamma$.)
If $L=1$, it will be a 2-D product torus, 
with the only difference that $p_-\rightarrow -p_-$ in the present construction.
If each Lagrangian surface corresponds to a state, i.e. $|\psi_-\rangle$
and $|\psi_+\rangle$, then we represent $|\Psi\rangle\!\rangle=|\psi_+\rangle\langle\psi_-|$
in the $|Q\rangle\!\rangle$ representation as
\be
\langle\!\langle Q|\Psi\rangle\!\rangle=\langle q_+|\psi_+\rangle\langle\psi_-|q_-\rangle.
\ee
Therefore, the semiclassical approximation is just a superposition of terms of the form
\be
\langle\!\langle Q|\Psi\rangle\!\rangle=A_J(Q) \exp [i S_J(Q)/\hbar], 
\ee
with
\be
A_J(Q)=A_{j-}(q_-)^*\>\> A_{j+}(q_+)
\ee
and
\be
S_J(Q)=\oint_0^Q P_J(Q')\cdot \rmd Q'.
\ee
Again this is in strict analogy to the construction of semiclassical
product states of higher degrees of freedom. Note that the projection
of the double Lagrangian torus onto $P$, or $Q$ is just the rectangle
discussed previously for product states, whereas the projections onto
the planes, $x_-$ and $x_+$ are specially singular.

The semiclassical approximation for a unitary operator, $\opU$,
that corresponds to a canonical transformation, $\C:x_-\rightarrow x_+$, 
has exactly the same form, i.e a superposition
\be
\langle\!\langle Q|U\rangle\!\rangle=\langle q_+|\opU|q_-\rangle=
U_J(Q) \exp [i S_J(Q)/\hbar],
\ee
for each branch of the function $P_J(Q)$ 
defined by the Lagrangian surface in double phase space.
Note that the situation with respect to projection singularities
is now reversed, as compared to $|\Psi\rangle\!\rangle$.
The fact that the projections of the Lagrangian surface, $\Lambda_\C$,
onto either $x_-$, or $x_+$ are both nonsingular in no way
guarantees that the projections onto the $P$, or the $Q$
Lagrangian planes will be likewise free of caustics.

Conversely, any function, $S(Q)$, is, at least locally, the generating
function of a canonical transformation through the implicit equations:
\begin{eqnarray}
\frac{\der S}{\der Q}=P(Q),\>\>
{\rm or}\>\>{\der S\over{\der q_+}} =p_+,\>\>{\der S\over{\der q-}}=-p_-.
\end{eqnarray}
Here we recognize the standard generating functions $S(q_-, q_+)$
in e.g. Goldstein \cite{Goldstein}.
If $S(Q)$ is quadratic, then these implicit equations will be linear,
so that the explicit transformation will result from a matrix inversion
(if it is nonsingular). There will be a single branch in $S(Q)$
for such a symplectic transformation
and it turns out that the semiclassical approximation is exact in this case.

The well known alternatives to these generating functions are usually
obtained by Legendre transforms. However, we can consider the $\pi/2$
rotation, $q_+\rightarrow p_+, p_+\rightarrow -q_+$, times the identity in $x_-$,
as an example of canonical transformation in double phase space: $X\rightarrow X'$.
Then $Q'=(q_-, p_+)$ is also a good Lagrangian plane which can be used
as the new coordinate plane  for the description of $\Lambda_\C$.
In the new coordinates the implicit equations for the canonical transformation
are just 
\begin{eqnarray}
\frac{\der S'}{\der Q'}=P'(Q'),\>\>
{\rm or}\>\>{\der S'\over{\der p_+}} =-q_+,\>\>{\der S'\over{\der q-}}=-p_-.
\end{eqnarray}
The correspondance with a semiclassical state,
\be
\langle\!\langle Q'|\Psi\rangle\!\rangle=A'_J(Q') \exp [i S'_J(Q')/\hbar], 
\ee
will be exact in the case of a symplectic transformation.
Note that $|Q'\rangle\!\rangle$ is a first example of an operator
basis that corresponds to a set of parallel Lagrangian planes 
in double phase space, which,
nonetheless, have internal coordinates that can be identified with 
a phase space on its own.

The crucial step is now to explore other kinds of
canonical transformations in double phase space \cite{AmHu80}.
In particular,
\be
Q'=\x=\frac{x_+ + x_-}{2}, \>\>P'=\y= \J(x_+ -x_-)=\J\vxi.
\label{centrechord}
\ee 
Here, the $\J$ symplectic matrix in single phase space is essential
to {\it canonize} what would be just a $\pi/4$ rotation. It accounts for the change
of sign in the $p_-$ coordinate. We will here have to bare the disconfortable situation
that the canonical coordinate in double phase space is $\y$,
but the geometrically meaningful variable in single phase space is $\vxi$,
the trajectory {\bf chord}, in the case of continuous evolution. The coordinate
$\x$ will be referred to as the {\bf centre}.

If we consider the {\it horizontal} Lagrangian planes $\y=constant$, 
each is identified with a uniform classical {\bf translation}.
Thus, we have departed from coordinate planes corresponding to dyadic operators
to those planes  in double phase that describe canonical transformations
and hence correspond to unitary transformations. In this case,
$x_-\rightarrow x_+=x_- +\vxi$ are the group of phase space translations, 
which include the the identity, i. e. the {\bf identity plane} is defined as $\vxi=0$.

On the other hand, the {\it vertical} plane, $\x=0$, defines the canonical
{\bf reflection} through the origin, $x_-\rightarrow x_+=-x_-$ (or inversion),
since all the chords for this transformation are centred on the origin.
Other vertical planes specify reflections through other points, 
$x_-\rightarrow x_+=-(x_- -2\x)$. The reflections do not form a group on their own
(no identity), but together with the translations 
they form the {\bf affine group} of geometry \cite{Coxeter} .

Since there is an exact correspondence between linear canonical transformations
and unitary transformations, each plane $\y=constant$ corresponds precisely to the
translation operator, $\opT_\vxi$, previously defined as \eref{transop}. Notice that this was
written with a phase that is a skew product involving $\vxi$, but we could
also use $\opT_\vxi= \exp[i\y\cdot\widehat x/\hbar]$. In terms of the previous
dyadic $|Q\rangle\!\rangle$ basis, this is expressed as 
\begin{equation}
\hat{T}_{\vxi}= \int \rm d q \left| q + \frac{\vxi_q}{2} 
\right\rangle \left\langle    q - \frac{\vxi_q}{2} \right|
                e^{i \vxi_p \cdot q/ \hbar}  \; ,
\label{optransq}
\end{equation}
a symmetrized Fourier transform (see e. g. \cite{AlmVal04}) .

Just as a $\pi/2$ rotation in single phase space, $q\rightarrow p, p\rightarrow -q$,
corresponds to a Fourier transform, so the transformation between {\it horizontal}
and {\it vertical} planes in double phase space is also achieved by a full Fourier transform
(except for an annoying factor of $2^L$):
\be
2^L\opR_{\x}=\int \frac{\rmd\vxi}{(2\pi\hbar)^L} \>\opT_{\vxi} \exp({i\over\hbar}\x\wedge\vxi).
\label{quantreflection}
\ee
In terms of the dyadic $|Q\rangle\!\rangle$ basis, we have
\begin{equation}
\label{quantrefleq}
2^L \hat{R}_\x = \int \rm d \vxi_q 
      \left| \mathbf q + \frac{\vxi_q}{2} \right\rangle 
\left\langle \mathbf q - \frac{\vxi_q}{2} \right|
       e^{i \mathbf p \cdot \vxi_q / \hbar},
\end{equation}
the complementary symmetrized Fourier transform to \eref{optransq}.

We are now free to switch from the usual (position) dyadic basis
to the unitary operator basis, $|\y\rangle\!\rangle=\opT_{\vxi}$: 
\be
\langle\!\langle \y|A\rangle\!\rangle= {\rm tr}\> \opT_{-\vxi}\opA=A(\vxi),
\ee
where we use $\opT_{-\vxi}=\opT_{\vxi}^{\dagger}$. 
$A(\vxi)$ is the {\bf chord representation} of the operator $\opA$. 
(Also referred to as the chord symbol). To verify that this is indeed the expansion
coefficient for an arbitrary operator in the basis of translation chords, we use
\be
{\rm tr}\> \opT_{\vxi}=(2\pi\hbar)^L \delta(\vxi) = \langle\!\langle \y|I\rangle\!\rangle
\ee
(note the double phase space analogy with $\langle p'|(p=0)\rangle=\delta(p')$ ),
as well as the quantum version of the group of translations:
\be
\opT_{\vxi_2}\opT_{\vxi_1} = \opT_{\vxi_1+\vxi_2} 
\exp\>[\frac{-i}{2\hbar} \vxi_1\wedge \vxi_2]
\ee    
(see e.g. \cite{Report}).
Then, the expansion
\be
\opA= \int \frac{\rmd\vxi}{(2\pi\hbar)^L} \>A(\vxi) \> \opT_{\vxi},
\ee
leads to
\begin{equation}
\fl \rm tr(\opT_{-\vxi}\widehat{A})=\rm tr \int \frac{\rm d\vxi '}{(2\pi \hbar )^L}\ 
A(\vxi ')\widehat{T}_{-\vxi}\widehat{T}_{\vxi '}=
\int \frac{\rm d\vxi '}{(2\pi \hbar )^L}\ A(\vxi ')
\exp \big[\frac{i}{2\hbar}\ \vxi '\wedge \vxi \big]
\>\rm tr\widehat{T}_{\vxi '-\vxi}=A(\vxi )\ .
\end{equation}
The chord representation is thus a second example of a representation of operators
in terms of an operator basis that can be identified uniquely to a phase space.
Indeed, each chord corresponds to a Lagrangian surface in 
double phase space and hence a particular uniform translation
in single phase space.

The next representation will be based on phase space reflections, $\opR_\x$.
But first it is worthwhile to examine some characteristcs of these operators.
Unlike the translations, they do not form a group on their own, though they combine
with the latter to form the affine group. 
The products are \cite{Report}
\be
\widehat{R}_\x \widehat{T}_{\vxi} =
\exp[ -\frac{i}{\hbar} \x \wedge \vxi] \>\> \widehat{R}_{\x-\vxi /2} ,
\ee
\be
\widehat{T}_{\vxi} \widehat{R}_\x  = 
\exp[ -\frac{i}{\hbar} \x \wedge \vxi ] \>\> \widehat{R}_{\x+ \vxi /2}
\ee
and
\be
\widehat{R}_{\x_2} \widehat{R}_{\x_1} =
\exp[ \frac{2i}{\hbar} \x_1 \wedge \x_2] \>\> \widehat{T}_{2(\x_2-\x_1)}.
\ee
Except for the phases, these are just the classical relations.
The last one is specially interesting. Note that $\opR_\x^2=\widehat I$,
the identity, hence the (degenerate) eigenvalues of $\opR_\x$ must
be either $+1$, or $-1$. Therefore these operators are Hermitian, as well as unitary.

Are they true observables? Consider the effect of $\opR_0$ 
on the eigenstates of the harmonic oscillator.
Taking $q\rightarrow -q$ and $p\rightarrow -p$, leads to a change of
sign for all the odd states, while preserving the even states.
In other words, the latter are just the $(+1)$-eigenstates, while 
the odd states are $(-1)$-eigenstates. Though it is hard to imagine
measuring the parity of a particle, we saw in section 2 that the parity decomposition
of even a classical wave can certainly be effected. 
Measurements of the eigenvalues of this {\it non-mechanical observable}
are currently performed for single photons 
in optical cavities \cite{Bertet02}.
It is true that these measurements are
performed on a mode of the electromagnetic field raher than a particle,
but it only makes sense to discuss the parity {\it within a specific mode} 
if it is quantized.

Reflection operators are very strange observables 
as far as phase space correspondence is concerned. It was discussed in
section 4 that usual observables correspond to smooth phase space functions
and their eigenvalues correspond to level curves if $L=1$.
This is just not the case of reflection operators with their
infinitely degenerate, $\pm 1$, eigenvalues.
In their dual role as both unitary and Hermitian (observable) operators,
reflections are almost schizophrenic: They are perfectly ordinary
unitary operators, corresponding to Lagrangian planes in double phase space,
but they do not correspond to any smooth classical function in phase space,
as expected of a mechanical observable.  

This should furnish sufficient motivation 
to investigate the representation of arbitrary operators in
terms of reflection centres. The assumption that 
\be
\opA= \int \rmd\x \>A(\x) \> 2^L\opR_{\x},
\label{Weylrep}
\ee
leads to
\be
\langle\!\langle \x|A\rangle\!\rangle= {\rm tr}\> (2^L\opR_{\x})\opA=
\rm tr \int \frac{\rm d \x'}{(2\pi\hbar)^L}A(\x') (2^L\opR_\x)(2^L\opR_{\x'})=A(\x).
\ee
This is the {\bf Weyl representation} of the operator $\opA$ (also known as the Weyl symbol).
Once again we use half the coordinates of double phase space, within a Lagrangian plane 
that is a phase space on its own, to describe a quantum operator. This perception
that we are really dealing with different phase spaces for each operator representation
was clearly stated in the excellent review by Balazs and Jennings \cite{BalazsJen} . 
What was lacking was merely the identification of each of these different phase spaces
with a specific foliation of Lagrangian planes in double phase space.

As far as unitary operators, $\opU$, are concerned, the semiclassical 
limit of the representations, either in terms of centres, or chords has exactly
the same form as for any other Lagrangian basis. For instance, the Weyl
symbol will be a superposition of terms, such as
\be
U(\x)= A(\x) \exp[iS(\x)/\hbar],
\label{USC}
\ee
in terms of the centre action, defined as
\be
S(\x)= \int_0^\x \y(\x')\cdot \rm d \x'=\int_0^\x \vxi(\x')\wedge \rm d \x'.
\ee
For symplectic transformations, the Lagrangian surface is a plane
and so there is only a single branch of the action function $S(\x)$, which is quadratic.
Then (\ref{USC}) is an exact representation 
of the corresponding quantum {\bf metaplectic transformation}.
However, in the general nonlinear case there may be caustics in the projection  of the
Lagrangian $\y(\x')$ surface onto the $\x$-plane.
Recall that this is just the plane that defines the identity operator, $\widehat I$
(corresponding to $\y=0$, or $S=0$). 

For the canonical transformation generated by a Hamiltonian, $H(x)$,
it turns out that the generating function has the limit \cite{Report}
\be
S_\epsilon (\x, t=\epsilon)\rightarrow -\epsilon H(\x)+\GO(\epsilon^3).
\ee
There are no caustics for small times
in the centre representation, 
since the corresponding Lagrangian surface is nearly horizontal.

The smooth real Hamiltonian itself can be equated to the Weyl symbol for the 
corresponding operator, $\opH$, within semiclassically small ordering terms.
This is the case of the Weyl representation for any observable that corresponds
classically to a smooth classical function of the points in phase space \cite{Vor76}.
Since we can always consider classical observables as infinitesimal generators of motion
through Hamilton's equations, it is appropriate to picture them as functions
on the $\y=0$ plane, so that the Hamiltonian vectors form a field on this plane
that indicates which way it will evolve. In contrast,
the chord symbol for these smooth mechanical observables is not at all smooth.
This is because the chord and centre symbols are related to each other 
through the Fourier transform,
\be
A(\vxi) = \frac{1}{\left(2\pi\hbar\right)^L} \int \rmd\x ~ 
\exp{\left(-\frac{i}{\hbar}\vxi\wedge\x\right)} A(\x),
\ee
just as the translation and reflection operators themselves in \eref{quantreflection}.
This Fourier transform takes the symbol for the identity, $I(\x)=1$, into
$I(\vxi)=\delta(\vxi)$ and a Taylor series in $\x$ into a
series of derivatives of $\delta$-functions. However,
we shall see in the next section that the chord representation of density operators
have very useful properties.

It is fitting to consider here another feature which distinguishes
the reflection operators from mechanical observables. Far from being
represented by a smooth phase space function, their centre representation
is just 
\be
R_\x(\x')=2^{-L}\delta(\x'-\x).
\label{weylrefl}
\ee
These singular functions
cannot be interpreted as corresponding to classical states 
(i.e. individual phase space points) because the $\opR_\x$ have the 
eigenvalue $-1$, so they are not density operators. 

Probably the first to remark on the general structure
of translations and reflections underlying the Weyl and the chord representations
were Grossmann and Huguenin \cite{GrossHug}.
There exists an exact correspondence, between these operators of the affine quantum group,
together with the unitary operators of the metaplectic group with the
classical transformations of the inhomogeneous symplectic group \cite{Vor76}.
In other words, all linear canonical transformations, including reflections and translations,
are exactly matched by quantum unitary transformations.
Thus, the unitary transformation, $\opU_{\bf C}$, corresponding to $x\rightarrow x'={\bf C} x$,
where {\bf C} is a symplectic matrix, takes
\be
\opR_\x \rightarrow \opR'_\x={\opU_{\bf C}}^\dagger\opR_\x\opU_{\bf C} =\opR_{\x'}\>\>\>
{\rm and} \>\>\>
\opT_\vxi \rightarrow \opT'_\vxi={\opU_{\bf C}}^\dagger\opT_\vxi\opU_{\bf C} =\opT_{\vxi'}.
\ee  
This has the consequence that
both the centre and the chord representations are invariant with respect
to metaplectic transformations, because the transformed operator, 
$\widehat A \rightarrow \widehat A'$ is represented by
\be
A'(\x)={\rm tr}\> {\opU_{\bf C}}\widehat A \; {\opU_{\bf C}}^\dagger\opR_\x
={\rm tr}\> \widehat A \; {\opU_{\bf C}}^\dagger\opR_\x {\opU_{\bf C}}
={\rm tr}\> \widehat A \;\opR_\x'=A(\x')
\ee
and, likewise, $A'(\vxi)=A(\vxi')$.

This section is concluded with some general formulae concerning
these representations.
For the trace of an operator, we have the alternative forms:
\be
\rm tr \>\opA= \rm tr\> \widehat I\> \opA=\langle\!\langle \opT_{\vxi=0}|A\rangle\!\rangle
             = A(\vxi=0)= \frac{1}{(2\pi\hbar)^L} \int \rm d \x\> A(\x).
\ee
The adjoint operator, $\opA^\dagger$, is represented by
\be
A^\dagger (\x)=[A(\x)]^*,\>\> \rm or \>\>\>A^\dagger(\vxi)=[A(-\vxi)]^*,
\ee
where $*$ denotes complex conjugation.
Thus, if $\opA$ is Hermitian, $A(\x)$ is real, though $A(\vxi)$
may well be complex. The Weyl or chord symbols for products of
operators is not at all obvious (see e.g. \cite{Report}), but
\begin{equation}
\rm tr\> \widehat{A}_2\widehat{A}_1 = \int \frac{\rm d\vxi}{(2\pi\hbar )^L} A_2 (\vxi )
A_1(-\vxi ) = \int \frac{\rm dx}{(2\pi\hbar )^L} A_2 (\x) A_1 (\x).
\label{traceprod}
\end{equation}

\section{The Wigner function and the chord function}

It is customary to alter the normalization of the centre and
the chord symbols for the density operator, $\widehat\rho$, so as
to define
\be
W(\x)= \frac {\rho(\x)}{(2\pi\hbar)^L} \rm\>\>\> and \>\>\>\>\>\chi(\vxi)=\frac {\rho(\vxi)}{(2\pi\hbar)^L},
\label{Wigner-chord}
\ee
respectively the {\bf Wigner function} and the {\bf chord function}.
Combining with the general definition of the Weyl representation \eref{Weylrep}
and the expression \eref{quantrefleq} for the reflection operator, 
we obtain the original definition of $W(\x)$, proposed by Wigner \cite{Wigner}. 
In both cases of \eref{Wigner-chord}, the representation of the trace of a product leads to
the expectation of any observable, $\opA$, as
\be
\langle\opA\rangle= \int \rmd\x \>W(\x) \>A(\x) = \int \rmd\vxi \>\chi(-\vxi) \>A(\vxi).
\ee
The first integral is more interesting because $A(\x)$ is at least
semiclassically close to the classical variable, which tempts us to
identify the Wigner function with a {\it nearly classical} probability
distribution. However, we will see bellow that $W(\x)$, though real
and normalized so
\be
\int \rmd\x \>W(\x) = 1,
\ee 
may well take on negative values. 

The chord function behaves like a classical characteristic function,
in as much as the {\bf moments} are
\begin{equation}
\label{moments}
\langle  q^n \rangle =  \rm tr\> \widehat{q}^n \>\widehat{\rho} = 
(i\hbar )^n \ \frac{\partial^n}{\partial \vxi_p^n} (2\pi\hbar)^L
\ \chi(\vxi) \Big|_{\vxi = 0} \; 
\end{equation}
and
\begin{equation}
\langle  p^n \rangle =  \rm tr\> \widehat{p}^n \>\widehat{\rho} = 
(-i\hbar )^n \ \frac{\partial^n}{\partial \vxi_q^n} (2\pi\hbar)^L
\ \chi(\vxi) \Big|_{\vxi = 0} \ . 
\end{equation}
Taking the zero'th moment, we obtain the normalization,
\be
1=  (2\pi\hbar)^L \>\>\chi (0),
\label{normchord}
\ee
because $\rm tr\>\oprho=\rho(\vxi=0)=1$.

Shifting the phase space origin to 
$\langle x\rangle=(\langle p\rangle,\langle q\rangle)$,
we can define the {\bf Schr\"odinger covariance matrix} \cite{Schro30}
just as its classical counterpart (\ref{Kov}), with
$\delta p^2=\langle \widehat{p}^2 \rangle$, $\delta {q}^2=\langle \widehat q^2 \rangle$
and $(\delta pq)^2=\langle (\widehat p \widehat q + \widehat q \widehat p)/ 2 \rangle$.
It is then obvious that the expansion of the chord function at the origin is just
\be
\chi(\vxi)=- \vxi\; {\bf K} \; \xi + ...
\ee
and we can interpret the {\bf uncertainty},
\be 
\Delta_{\bf K}= \sqrt{\det {\bf K}}, 
\ee
as proportional to the volume of the ellipsoid: $\vxi\; {\bf K} \; \xi=1$.
Evidently, this volume is invariant with respect to symplectic transformations,
so that $\Delta_{\bf K}$ is a symplectically invariant measure of the uncertainty
of the state.

The projection of the Wigner function,
\be
\int \rmd \mathbf p\>\> W(\mathbf p,\mathbf q) ={\rm Pr} (q)
\label{qproj}
\ee
is a true probability for position measurements \cite{Wigner}. 
Furthermore, the invariance of the chord and
the centre representations with respect to symplectic transformations
then guarantees that the projection of the Wigner function along any
set of Lagrangian planes $\mathbf  p'$ supplies the probability distribution
for the conjugate variable $\mathbf  q'$. In particular, the 
probability ${\rm Pr}(p)$ results from the projection of $W(\mathbf p,\mathbf q)$ 
with respect to $\mathbf q$. All these planes are Lagrangian, so it follows that
the projection of the Wigner function onto any Lagrangian plane in phase
space is a probability distribution for the corresponding variable.
It may appear somewhat contrived, as far as measurement is concerned, to consider general
linear combinations of position and momentum.
However, it should be recalled that
these observables will evolve from an initial position for the motion driven by
any quadratic Hamiltonian, even including free motion through a laboratory. 
The reconstruction of the Wigner function 
from a suitable set of these marginal distributions is known as quantum tomography. 
This is achieved through the {\it Radon transform} (see e. g. \cite{Radon}).

It is equally remarkable, but less well known, that the characteristic function
corresponding to the marginal probability distribution for positions is obtained by
merely taking a section of the chord function:
\be
\int \rmd \mathbf q \>\>{\rm Pr}(\mathbf q)\>\exp{\left(-\frac{i}{\hbar}\eta_q\cdot \mathbf q \right)} 
= (2\pi\hbar)^L \>\>\chi (0,\eta_q).
\ee
Since, the chord function is also symplectically invariant, it follows that
the characteristic functions for all the probability distributions, which
result from Wigner projections onto Lagrangian planes, are equal 
to the corresponding sections of the chord function.

So far we have emphasised the seemingly classical aspects 
of the Wigner function. However, it must be remembered that
the Weyl representation is defined in terms of a very anomolous observable,
as far as classical correspondence is concerned.
In order to reveal the full quantum nature of the Wigner function,
let us divide the Hilbert space of quantum states into even and odd
subspaces for a given reflection operator, $\opR_\x$.
This is achieved through the projection operator introduced by
Grossmann \cite{Grossmann} and Royer \cite{Royer},
\begin{equation}
\widehat{P}^\x_\pm = \frac{1}{2}
\left( 1 \pm \widehat{R}_\x \right) \; ,
\label{projector}
\end{equation}
so that, in its turn, we can express each reflection operator
as the superposition of this pair of projections onto the even and
the odd subspaces:
\be
\opR_\x= \widehat{P}^\x_+ -\widehat{P}^\x_-.
\ee
But 
\be
\rm tr\> \widehat \rho\>\widehat{P}^\x_\pm =\rm Pr_\pm^\x
\ee
is just the probability
of measuring $\opR_\x$ to have the eigenvalue $\pm 1$, so it follows that \cite{Royer}
\be
W(\x)={1\over (\pi\hbar)^L}[\rm Pr_+^\x - \rm Pr_-^\x]= 
{1\over (\pi\hbar)^L}[2\rm Pr_+^\x-1].
\ee
We thus find that the Wigner function does not admit the interpretation
as a probability distribution in phase space, because it can certainly be negative.
Even so, it is a simple linear function of a distribution of probabilities of
positive eigenvalues for all possible reflection measurements.
Its maximum possible value $(\pi\hbar)^{-L}$ is attained for any point, $\x$,
such that $\widehat{P}^\x_+\> \widehat \rho=\widehat \rho$, whereas the commutation 
of the density operator with $\widehat{P}^\x_-$ specifies a phase space point
where $W(\x)=- (\pi\hbar)^{-L}$.

Let us now investigate the effect of reflections and translations on a
density operator. Evidently, the centre and chord representations
are specially suitable for this purpose. In the case of a phase space
translation by the vector, $\eta$, i.e. $\widehat{\rho}_{\veta}=
\widehat{T}_{\veta}\>\widehat{\rho}\>\widehat{T}_{-\veta}$,
the respective Wigner and chord functions become
\be
W_{\veta}(\x)=W(\x-\veta ) \>\>\> \rm and \>\>\> 
\chi_{\veta}(\vxi ) = e^{i \eta \wedge \vxi / \hbar}\>\> \chi(\vxi ),
\label{translate}
\ee
which shows that, unlike the Wigner function, the chord function is not generally real.
The sensitivity of a state to translations is described by the {\bf phase space correlations} 
of a given density operator, defined as \cite{AlmVal04} 
\begin{equation}
\fl C(\vxi)=
\rm tr \> \widehat{\rho} \>\widehat{T_\vxi}\>\widehat{\rho}\>\widehat{T}^\dagger_\vxi=
(2\pi \hbar )^L \int \rm d\x \>W(\x)\> W(\x-\vxi)=
(2\pi \hbar )^L \int \rm d\veta \; e^{ i \veta \wedge \vxi / \hbar}
\left| \chi(\veta )\right|^2 \; .
\label{pscor}
\end{equation}
>From the reciprocal relation that supplies the intensity of the chord function
as the Fourier transform of these correlations and the normalization condition
\eref{normchord}, we see that
\be
\int \rm d\veta \;C(\veta)=(2\pi \hbar )^{3L} |\chi(\veta=0)|^2=(2\pi \hbar )^L. 
\ee
So, even though these correlations are defined in terms of classical translatons 
in phase space, they are purely quantum and disapear in the classical limit.
However, if we fix $\hbar$ and adopt this constant as our phase space scale,
then we can picture $C(\vxi)$ as a classical-like phase space distribution for which the
characteristic function is just $|\chi(\vxi)|^2$.
 
Specializing to the case of a pure state,
$\widehat{\rho}=| \psi \rangle \langle  \psi | $,
we find that
\begin{equation}
\langle \psi |\opT_\vxi|\psi \rangle\ = (2\pi \hbar )^L \chi(-\vxi )\ ,
\end{equation}
so that the phase space correllations take the form \cite{AlmVal04} 
\be  
 C(\vxi)= |\langle \psi |\opT_\vxi|\psi \rangle|^2 = (2\pi\hbar)^{2L}|\chi(\vxi )|^2 .
\label{purecor} 
\ee
Thus, for instance, in the case that $\vxi=(0,\vxi_q)$, 
the phase space correlations, $C(\vxi)$,
are just the usual spacial correlations inferred from
neutron scattering experiments. Nonetheless, we must be carefull to distinguish
between phase space correlations and the correlations between the
quantum measurements of observables defined on the different components
of a bipartite system, such as enter the CHSH inequality.
For a pure state, (\ref{purecor}) is the square modulus of the expectation 
for a translation, which is not a quantum observable. However, 
(\ref{pscor}) defines the phase space correlation 
in the same way as for a classical  distribution.

The chord function always assumes its maximum
value $1/(2\pi \hbar)^L$ at the origin. But also an average of overlaps 
cannot exceed one, so $\chi(0)$ is the maximum even for mixed states.
As for the correlations, we always have 
\be
{\rm tr}\>\oprho^2= (2\pi\hbar)^{L}\int \rmd\x \>[W(\x)]^2 
=(2\pi\hbar)^{L} \int{\rmd\vxi} \> |\chi(\vxi)|^2= C(0),
\ee
being that ${\rm tr}\>\oprho^2 =1$ for pure states. 
But consider a mixture of orthogonal states,
\be
\oprho= \sum_j {\rm Pr}(n)\> |n\rangle\langle n|,
\ee
then the {\bf purity}
\be
C(0)=\sum_j {{\rm Pr}(n)}^2\leq 1.
\label{purity1}
\ee
Another form in which this quantity appears is 
the {\bf linear entropy} : $1-\rm tr \>\widehat{\rho}^2$.
This may be considered as a first order expansion of the {\bf von Neumann entropy}:
\be
-{\rm tr}\>\oprho\>\ln\oprho=-\int \rmd\x \>W(\x)\>\ln[(2\pi\hbar)^{L}\>W(\x)], 
\label{vonneumann}
\ee
a quantum version of the classical Shannon entropy.
In \cite{AlmVal04} the correlations were normalized by the purity so as to be always unity
a the origin, but it is convenient to include 
this quantity as a special case of the correlations.

General invariance with respect to Fourier transformation characterizes
the correlation in the case of pure states. 
Indeed, inserting the above expression in the definition of the
phase space correlation \eref{pscor}, we obtain \cite{AlmVal04} 
\begin{equation}
\label{Finvariance}
 C(\vxi)=\int \frac{ d \eta}{(2\pi\hbar)^L} \;
e^{i \veta \wedge \vxi / \hbar}\> C(\eta)  \; .
\end{equation}
This is a remarkable property of all pure states and is in no way
restricted by special symmetry properties that will be shown to 
relate certain Wigner functions to their respective chord functions.
An immediate consequence is that oscillations of the phase space correlation 
of a pure state involving a large displacement, $\vxi$, are necessarily
bound to small ripples on the scale, $|\vxi|^{-1}$, in the direction, $\J\vxi$.
Of course, these small scale oscillations of the phase space correlations, 
which have been attractively described as {\it subplanckian} \cite{Zurek},
show up in the pure state Wigner function because of (\ref{pscor}). 

The Fourier invariance condition (\ref{Finvariance}) includes as a special case 
the more familiar one obtained by tracing over the full pure state condition 
$\hat{\rho}^2=\widehat{\rho}$. 
It follows that the difference of both sides of (\ref{Finvariance}) for 
each chord $\vxi$ generalizes (\ref{purity1}) as a measure 
of the degree of {\it purity} of a state. 
All the same, the loss of the phase information in $C(\vxi)$,
but contained in the chord 
function, would seem to imply that these are necessary 
conditions, whereas the full sufficient condition of purity is 
$\hat{\rho}^2=\widehat{\rho}$, 
which is expressed in the chord representation as \cite{Report}
\begin{equation}
\int \rm d\veta \> \chi(\veta )\> \chi(\vxi -\veta ) \>
e^{i \vxi \wedge \veta /2\hbar} = 
\int \rm d\veta \> \chi_{\vxi/2}(\veta ) \> \chi(\vxi -\veta ) = 
\chi(\vxi ) \; ,
\end{equation}
with $\chi_{\vxi/2}(\veta )$ defined by (\ref{translate}).
\footnote{For distributions, ${\rm Pr}(n)$, over eigenstates $|n\rangle$ of an observable with discrete spectrum,
the condition ${\rm Pr}(n)^2={\rm Pr}(n)$ also singles out a {\it pure state},
${\rm Pr}(n)=\delta_{n,m}$, but this condition is not generalizable to a continuous spectrum.}
However, the particular condition $C(0)=1$ is indeed a suficient condition,
because, for any mixture of pure states, $\oprho=\sum_n {\rm Pr}(n)| \psi_n \rangle \langle  \psi_n | $,
we obtain
\be
{\rm tr}\>\oprho^2= 1 - \sum_{n \neq n'} {\rm Pr}(n){\rm Pr}(n')[1 - |\langle \psi_n |\psi_{n'} \rangle|^2]. 
\ee

A single phase space point does not correspond to any pure state in Hilbert space. 
The only pure states that are {\it classical-like}, i.e. have positive Wigner functions,
are either coherent states, or their immage by a symplectic transformation \cite{Hudson, Tatarskii}.

Let us now consider the effect of measuring a general phase space reflection,
$\opR_\x$. The density operator, $\oprho$, will be projected 
by $\widehat{P}^\x_\pm$, defined by \eref{projector}, onto either
the even, or odd subspace for this particular relection:
\begin{equation}
\widehat{\rho}^\x_{\pm} = 
\frac{\widehat{P}^\x_{\pm} \>\widehat{\rho}\>\widehat{P}^\x_{\pm}}
{   \rm tr\> \widehat{\rho} \>\widehat{P}^\x_{\pm}} \; .
\end{equation}
The Weyl symbol for $\widehat{\rho} \>\widehat{P}^\x_{\pm}$
coincides with the symmetric Wigner function, $W_\x^{\pm}(\x')$,
within a normalization factor, so that we obtain \cite{AlmBro04}:
\begin{equation}
\fl W_\x^{\pm}(\x') = 
(\pi\hbar)^{-L} \rm tr \widehat{R}_{\x'} \left(\frac{1\pm\widehat{R}_\x}{2}\right)\widehat{\rho}^\x_{\pm} 
= \frac{W_\x^{\pm}(\x')}{2}
\pm\frac{2^L e^{ 2i \ \x' \wedge \x / \hbar } \rm tr \>\widehat{T}_{2(\x'-\x)}\widehat{\rho}^\x_{\pm}}{2}\> \; .
\end{equation}
It follows that the Wigner function and the chord function for a
reflection symmetric density operator are trivially related.
Shifting the origin of phase space to the symmetry point leads to \cite{AlmVal04}
\begin{equation}
W^{\pm}_0(\x)=\pm \ 2^L\chi^{\pm}_0(-2\x) \; .
\label{symwig}
\end{equation}
Thus, all Wigner functions for density operators that commute with
a reflection symmetry attain the largest amplitude at the symmetry point,
but this will be negative in the case of odd symmetry.
 
Let us consider some standard examples of Wigner and chord functions.
All the following cases are related to eigenstates of a harmonic oscillator 
with one degree of freedom and unit mass.

i) Coherent states: the Wigner
function is just a Gaussian centered on $\eta$,
\begin{equation}
\label{wcoherent}
\fl W_{\eta}(\x) =
 \frac{1}{\pi \hbar}
 \exp \left[
-\frac{\omega}{\hbar}    \ \left(\mathbf  q-\eta_q\right)^2 -
 \frac{1}{\hbar \omega}  \ \left(\mathbf p-\eta_p\right)^2      
     \right] 
 \stackrel{\omega=1}{\longrightarrow}
 \frac{1}{\pi \hbar} e^{-(\x-\eta)^2/\hbar} \; ,
\end{equation}
whereas,
\begin{equation}
\fl \chi_{\eta}(\xi )= 
\frac{1}{2\pi \hbar}
 \exp 
 \left(\frac{i \eta \wedge \xi}{\hbar}\right)
 \exp \left[-\frac{\omega}{\hbar}
 \left(\frac{\xi_q}{2}\right)^2-\frac{1}{\hbar\omega}
 \left(\frac{\xi_p}{2}\right)^2\right] 
\stackrel{\omega=1}{\longrightarrow}
\frac{1}{2\pi \hbar}
e^{i \eta \wedge \xi / \hbar}
e^{-\xi^2 / 4\hbar} \; .
\end{equation}
So, any translation of the coherent state merely alters 
the phase of the Gausssian chord function that sits on the origin.
The coherent states, or more generally all equivalent
Gaussian states obtained from them by symplectic transformations,
are the only examples of pure states for which the Wigner function
is nowhere negative \cite{Hudson}. This is one of the reasons why these are sometimes
considered to be the most classical of pure quantum states.
Since the projection of a Gaussian is also a Gaussian, 
the measurement of position, or any other Lagrangian phase space coordinate,
does not display interference fringes.
The fact that the uncertainty, $\Delta=\delta p\delta q=\hbar$, 
is minimal allows us to interpret them as {\it quantum phase space points}.

\begin{figure}
\includegraphics[width=15cm]{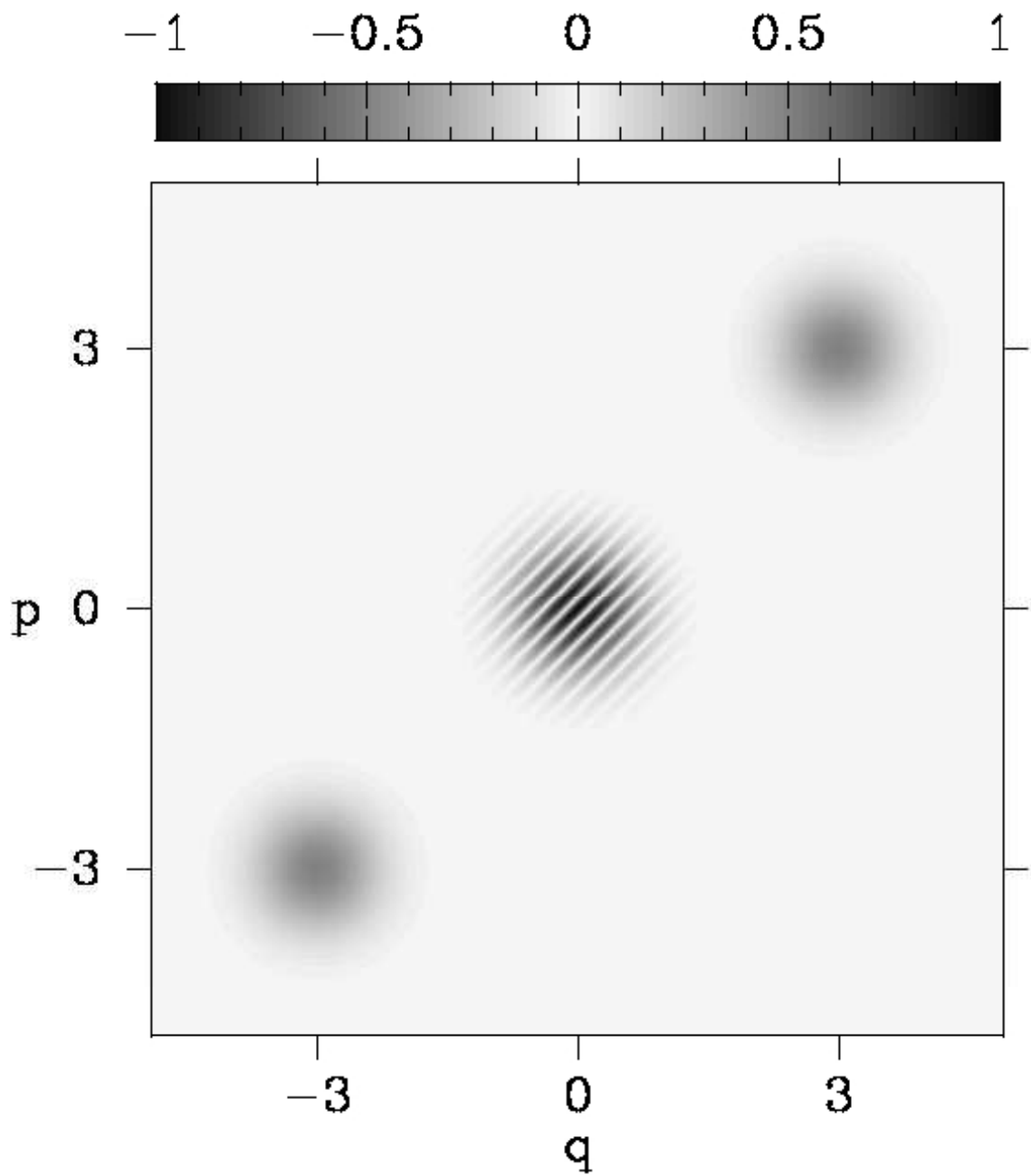}
\caption{The Wigner function for the Schr\"{o}dinger cat state 
displays a pair of {\it classical} Gaussians, one for each coherent state,
and a third Gaussian modulated by interference fringes halfway between them.
The chord function is a mere rescaling of the Wigner function if the midpoint
lies on the origin.}
\label{cat}
\end{figure}

(ii) A superposition of a pair of coherent states, 
$|\eta \rangle \pm |- \eta \rangle$ 
is sometimes known as a {\it Schr\"{o}dinger cat state}. 
Its Wigner function is \footnote{Here and below we set $\omega =1$.}
\begin{equation}
\fl W_{\pm}(\x) = 
\frac{1}{2\pi \hbar \,(1\pm e^{-\eta^2/\hbar})}
\left[e^{-(\x-\eta )^2 / \hbar } +
      e^{-(\x+\eta )^2 / \hbar } \pm 
    2 e^{-\x^2         / \hbar } \cos \frac{2}{\hbar} \x\wedge \eta \right] \; .
\label{Wigcat}
\end{equation}
It consists of two {\it classical} gaussians centred on $\pm \eta$ and 
an interference pattern with a gaussian envelope centred on their 
midpoint.  
The frequency of this oscillation increases with the separation 
$|2 \eta |$. In fig.\ref{cat} the displacements, $\pm\eta$ have been chosen 
as $(\pm 3, \pm 3)$. 
The phase of the pair of coherent states merely shifts the phase of the
interference fringes, so that the midpoint is an absolute maximum for $W_+(\x)$
and an absolute minimum for $W_-(\x)$.
It might be supposed that, for small $\eta\rightarrow 0$,
we would have $W_+(\x)>0$ for all $\x$, but it is 
easy to verify that there are very shallow negative regions
far removed from the classical superposed Gaussians, 
in agreement with \cite{Hudson, Tatarskii}.  
The interference pattern of the Wigner function does not
survive the projection orthogonal to $\eta$: In this direction, the interference disappears
to produce a purely classical pattern. Conversely, the projection along $\eta$ 
is marked by interference fringes.

For the chord function,
\begin{equation}
\fl \chi_{\pm}(\xi ) = 
\frac{1}{4\pi \hbar \,(1\pm e^{-\eta^2/\hbar})}
\left[e^{-(\xi/2-\eta )^2 / \hbar } +
      e^{-(\xi/2+\eta )^2 / \hbar } \pm 
    2 e^{-\xi^2       / 4\hbar } 
\cos \frac{1}{\hbar} \xi \wedge \eta \right] \; ,
\end{equation}
this same configuration has to be reinterpreted. 
Now the local phase space correlations of the individual coherent states, as in (i),
are placed in the neighbourhood of the origin, where they interfere, 
while their cross-correlation generates new Gaussians centred on 
the separation vectors $\pm 2\eta$. 
The general case of coherent states 
$|\eta_1 \rangle$ and $|\eta_2\rangle$ 
merely leads to Gaussians centred on
$\pm (\eta_1-\eta_2)$ with addition of the phase factor 
$\exp [i (\eta_1+\eta_2) \wedge \xi  /2\hbar]$.

Recalling that the phase space correlations of a pure state
are just the square modulus of the chord function, we can
immediately verify the general relation betweeen
large and small scale structures \eref {Finvariance} 
in the case of Schr\"{o}dinger cat states.
Indeed, the spacial freequency of the oscilations of the
chord function increases directly with the separation of the
pair of coherent states.

The particular superpositions of coherent states,
$|+\rangle$ and $|-\rangle$, are respectively even and odd
eigenstates of the parity operator $\opR_0$, i.e. reflection about the origin.
Therefore, they are the two possible states that could be produced by a 
parity measurement effected on the single coherent state $|\eta\rangle$.
Thus, the parity measurement would generate a sizable probability of finding
a particle near $x=-\eta$, even though this were most unlikely before the
measurement.

The states $|\pm\rangle$ are orthogonal, even though the coherent states, 
$|\eta\rangle$ and $|-\eta\rangle$, are not. 
It is true that such a pair of coherent states will be nearly orthogonal
if $\eta$ is large enough and thus considered to form a qubit.
Within this approximation, 
the symmetrical states would then be a mere unitary transformation of 
a single qubit. However, no approximation is needed in this process
of carving a qubit from an infinite dimensional system, if we use $|\pm\rangle$
as the original basis states. We would then consider a {\it common garden} coherent state
to be the superposition of a symmetrical pair of Schr\"odinger cats.
(Is there some approximation involved?) Indeed, this generation of a qubit
by a reflection is not limited to coherent states, but could in principle be
realized for any unsymmetrical initial state.  

\begin{figure}
\includegraphics[width=15cm]{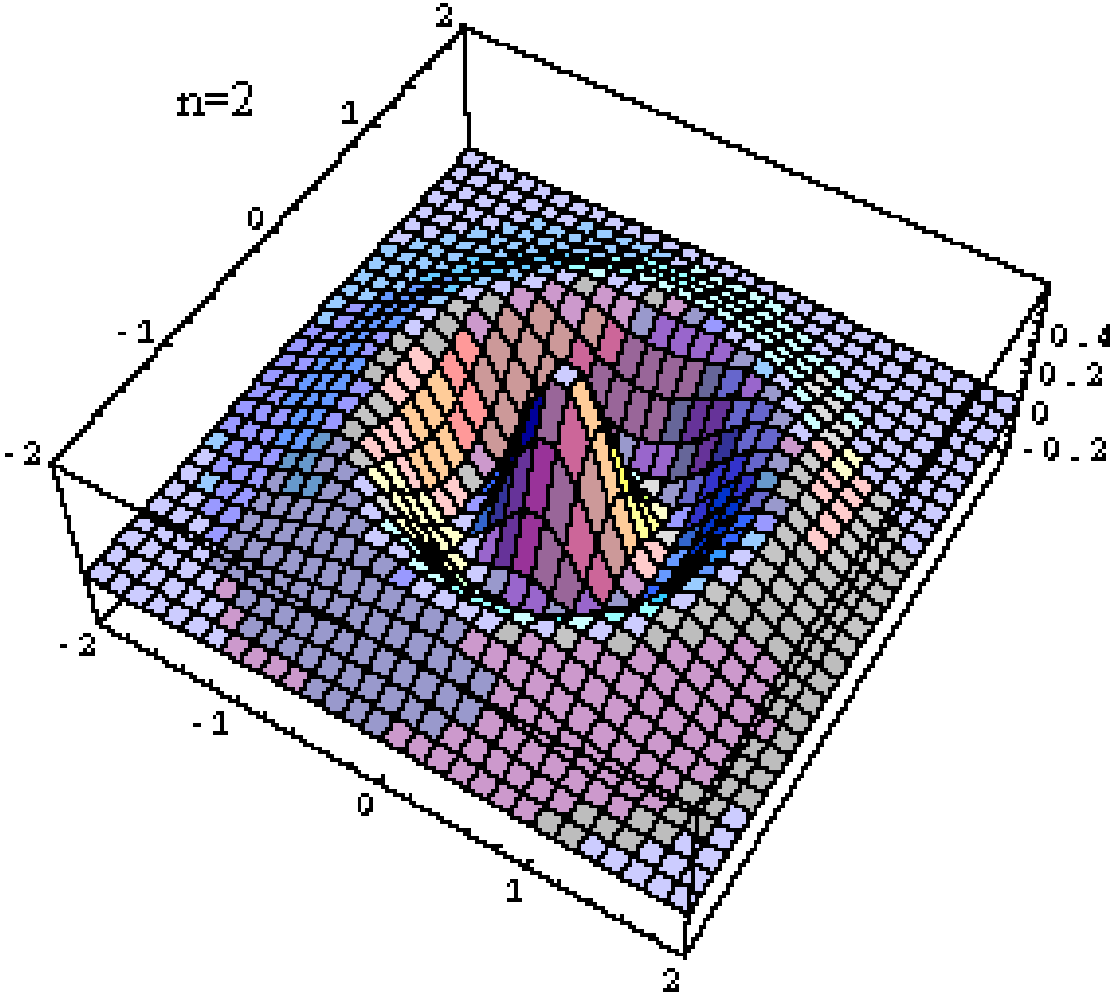}
\caption{The Wigner function for the $n=2$ Fock state. 
The classical Bohr-quantized circle 
lies just outside the maximum of the outer fringe.
The chord function is just a rescaling of the Wigner function.}
\label{Fock}
\end{figure}

(iii) Fock states, $|n \rangle$, i.e., the excited states of 
the harmonic oscillator, also have reflection symmetry with respect 
to the origin. 
Thus, from the exact Wigner function, first derived by
Gr\"onewold \cite{Gronewold46},
\begin{equation}
\label{Fockst}
W_n(\x) = 
\frac{(-1)^n}{\pi \hbar} e^{-\x^2/\hbar}
L_n \left(\frac{2\x^2}{\hbar}\right) \; ,
\end{equation}
where $L_n$ is a Laguerre polynomial, we obtain the chord function
\begin{equation}
\chi_n (\xi ) = \frac{e^{-\xi^2/4\hbar }}{2\pi \hbar}
L_n \left( \frac{\xi^2}{2\hbar} \right) \; .
\end{equation}
It is interesting to note that the symmetry centre, which produces
the maximum amplitude of the Wigner function, is nowhere near the
classical  manifold with energy 
$E_n=\left(n+\frac{1}{2}\right) \hbar\omega$. 
However, this point lies in a region of narrow oscillations, 
so that it does not affect the average of smooth observables.
Fig.\ref{Fock} shows the Wigner function for the Fock state with $n=2$;
the origin is a maximum because of the posititive parity. 
The unfolding of this peak for nonsymmetric Wigner functions
is discussed in section 10.

The Wigner function exhibits the interference fringes for
the measurement of any variable $ap+bq$. In the case of the Fock
state, these are always present. A simple way to see this is
that any direction for the projection will be somewhere tangent 
to each of the continuous curves that form the Wigner function fringes.
These regions dominate the projection. This example thus illustrates
the {\it necessity} for the Wigner function to have negative regions:
This is the only way that interference can result from a mere
projection in phase space. 

The Fock states are an example of a complete parity basis,
which is even or odd according to the state label, $n$.
Hence, if a pure state, $|\psi\rangle$, is specified in this basis,
then 
\be
W_\psi(0)={1\over{\pi\hbar}}\sum_n [|\langle 2n|\psi\rangle|^2-|\langle (2n+1)|\psi\rangle|^2].
\ee
Such a decomposition can in principle be achieved for general arguments
of the Wigner function, but then it is necessary to translate the whole
Fock state basis instead of just the ground state, as in the definition of coherent states.
If we similarly translate the Hamiltonian, it will commute with $\opR_\x$
instead of commuting with $\opR_0$. The eigenstates of all such Hamiltonians 
will form a good odd-even basis. The difficulty with defining a semiclassical
correspondence for both these classes of eigenstates is that the odd and even
Bohr-quantized curves approach each other without a limit as $\hbar\rightarrow 0$. 

All the above examples are singled out by some point of reflection
symmetry, which needs to be chosen as the origin for the chord
function to be real. The chord function must assume its maximum
value $1/(2\pi \hbar)^L$ at the origin, whatever the symmetry, 
because of normalization.
The Wigner amplitude, $|W(\x)|$, need not have such a prominent peak 
in general. 
However we shall see in section 10 that the large scale 
features of the semiclassical forms of the Wigner function and the 
chord function maintain a mutual correspondence, even in the absence 
of a reflection symmetry. 

It is important to note that the commutation,
$[\oprho, \opR_0]=0$, guarantees that $W(\x)$ is a symmetric function
with respect to (classical) reflection at the origin.
This is a consequence of the fact that, if $\oprho_R= \opR\>\oprho\>\opR$,
then $W_R(\x)=W(R(\x))$, the classical reflection of the argument. 
However, it is the maximum (or minimum) value at the origin which guarantees
that the density operator is pure with respect to parity, i.e. it is either
$\oprho_+$, or $\oprho_-$. Indeed, even though a mixture of an even density 
and an odd density (i.e. $\oprho=c_+ \oprho_+ +c_-\oprho_-$)
will trivially satisfy $[\oprho, \opR_0]=0$, we see that 
$W(0)=(c_+ -c_-)/(\pi\hbar)^L$ will not be maximal.

Perhaps the converse property is of even more interest: 
If a Wigner function is symmetric about the origin,
but $|W(0)|<(\pi\hbar)^{-L}$, then the state must be a mixture.
After all, if it is a mixture of parities, it cannot be a pure state.
It is only if the mixture is restricted to states of the same parity
that it will not be detected by $W(0)$.
There are many measures of degrees of mixedness, or impurity,
but it is specially nice to be able to spot this property by a 
mere glance at the Wigner function. Furthermore, the Wigner function,
i.e. the parity decomposition, is a measurable property \cite{Englert, LutterbachDav, Bertet02}. 

The Wigner function may be considered as a field of probabilities for parity
decompositions in phase space. Each reflection separates the
infinite dimensional Hilbert space into a pair of orthogonal components.
If we just consider a single reflection, this goes a long way to reducing
the Hilbert space to that of a single qubit, a two state system.
No matter how classical the appearance of the Wigner function 
(i.e. it may be smooth and positive) it is always fully quantum as far as
parity measurements are concerned. The situation is quite different,
for instance, for  position measurements. Then there is an important
difference between the Wigner function for a pure Schr\"odinger cat state
and a mixture of cats with different phases. This is revealed by the fine interference
fringes between the two classical regions, but is even more clearly displayed
by the pair of correlation peaks far from the origin of the chord function.
The relation between these features and entanglement is discussed in section 8.

There is a vast litterature concerning the Wigner function.
Only a few topics have been mentioned here and it has been necessary
to leave out even as relevant a topic as quantum tomography.
The adaptation of the Wigner function for finite Hilbert spaces 
is of special relevance for quantum computing and quantum information theory.
Then the rule that each quantum state corresponds to a volume of $(2\pi\hbar)^L$
in classical phase space, restricts the overall phase space volume. Thus, one must first
face a choice of the topology in which to compactify phase space.
It turns out that the simplest choice is a torus, though single qubits
are more naturally displayed on a {\it Bloch sphere}.
In spite of the intrinsic interest in many of the aspects of
finite space Wigner functions \cite{Wootters,MiqPazSar}, there is no overall agreement
on the choice of Wigner function propertise to emphasise.
Not all formalisms lead to a corresponding natural definition of
a conjugate chord function as in \cite{RivAlm}, nor is there an overall preocupation
with invariance with respect to those symplectic transformations
which preserve the torus topology of phase space \cite{RivSarOA}. 
A final difficulty concerns the appearance of ghost immages 
and dimensionally dependent features \cite{ArgDit}.  

So far nothing has been said of an alternative phase space representation,
the \textsl{Husimi fuction}\cite{Husimi, Takashi}. 
Defined in terms of coherent states $|\eta\rangle$ as
\be
\rho_H(\eta)= \langle\eta|\rho|\eta\rangle = {\rm tr}\>\rho\>|\eta\rangle\langle\eta|,
\ee
it can be interpreted as a smoothed Wigner function, 
\be
\rho_H(\eta)=\int \rm d\x\>W_\eta(\x)\>W(\x),
\label{Husimi}
\ee
because of (\ref{traceprod}).
The lack of purity of a state can be described in terms of the {\bf Wehrl entropy}, 
\be
S_W= -(2\pi\hbar)^L \int {\rm d}\eta \>\rho_H(\eta) \ln \rho_H(\eta).
\label{Wherl}
\ee
According to Wehrl's inequality \cite{Wehrl}
(see also \cite{BengZycs}), the Wehrl entropy is always bounded from below
by the von Neumann entropy (\ref{vonneumann}).
For more recent developments concerning the
Wehrl entropy, see e.g. \cite{MinZicz}.

The Husimi function is most appropriate for the study of {\it quantum chaos},
because it highlights the classical region. But such a downplay
of the quantum interferences, achieved by coarse graining
the Wigner function, is not what one would ordinarily seek in
quantum information theory. In a way, this is just the opposite
of the chord function, which squashes all classical structure to
the neighbourhood of the origin, so as to display the purely 
quantum coherences. It is remarkable that both these antithetical
representations are intimately related to the translation operators,
since the Husimi fuction for a pure state,
$\widehat{\rho}=| \psi \rangle \langle  \psi |$, can be rewritten as 
\be
\rho_H(\eta)=|\langle \psi|\eta\rangle|^2
=|\langle \psi|\opT_\eta|(\eta=0)\rangle|^2.
\ee
Hence, the basic difference with respect to $C_\eta$ in \eref{purecor}
is the exchange of $|0\rangle$, 
the Gaussian ground state of the harmonic oscillator, for $|\psi\rangle$ itself. 

A further comment is that the quantum interferences
are displayed by the isolated zeroes of Husimi functions \cite{LebVoros}, 
in the case that $L=1$.  A uniform distribution of zeroes 
has been used to characterize the eigenstates of classically chaotic systems. 
Even though this is of great theoretical interest, these zeroes 
are ususally located in regions where the Husimi function is already tiny, so
that they may be very hard to compute. For instance, 
in the case of the cat state (\ref{Wigcat}) with small $\eta$, 
they are found in the shallow negative regions where $|W_+(\x)|$
is exponentially small.

\section{The partial trace: sections and projections}

Recall that the representation of operators, $\opA=|A\rangle\!\rangle$,
in a given basis, such as $\langle\!\langle Q|A\rangle\!\rangle$,
corresponds to the  folliation of the double phase space, $X=(P,Q)$,
by a set of Lagrangian planes, $Q=constant$. Performing linear canonical
transformations in double phase space, we are free to choose 
the alternative coordinate planes, $Q=(q_-, q_+)$,
or $Q=\x$, or $Q=\y=J\vxi$ among others. In all cases, it is the fact that
\be 
\langle\!\langle Q'|Q\rangle\!\rangle=\delta(Q'-Q)
\ee
which permits us to identify the expansion coefficient in
\be
\opA= |A\rangle\!\rangle=\int \rmd Q \>A(Q) \> |Q\rangle\!\rangle
\ee
with $\langle\!\langle Q|A\rangle\!\rangle$.
 
Let us now assume that the phase space is a product of
a pair of phase spaces, $X=X_1\otimes X_2$, each with $2L_j$
dimensions, and that these correspond to Hilbert spaces, ${\bi H}_j$,
so that ${\bi H}={\bi H}_1\otimes {\bi H}_2$. Then we can always decompose the 
Lagrangian planes chosen as a basis for double phase space 
as the product $Q=Q_1\otimes Q_2$, corresponding to, operators
$|Q\rangle\!\rangle=|Q_1\rangle\!\rangle\otimes |Q_2\rangle\!\rangle$. 
Thus the complete $|Q\rangle\!\rangle$ representation becomes
\be
\opA= |A\rangle\!\rangle=
\int \rmd Q_1 \rmd Q_2 \>A(Q_1, Q_2)\>|Q_1\rangle\!\rangle\otimes |Q_2\rangle\!\rangle.
\ee

The definition of the partial trace is then
\be
\rm tr_2\>\opA=\rm tr_2\> \opI_2\>\opA=
\int \rmd Q_1 \rmd Q_2 \>A(Q_1, Q_2)\>
|Q_1\rangle\!\rangle\>\langle\!\langle I_2 |Q_2\rangle\!\rangle,
\ee
so that 
\be
A_1(Q_1)= \int \rmd Q_2 \>A(Q_1, Q_2)\langle\!\langle I_2 |Q_2\rangle\!\rangle
\ee
defines the $|Q_1\rangle\!\rangle$ representation of a reduced operator $\opA_1$,
which  acts on the Hilbert space ${\bi H}_1$. It is well known that in the case
of the density operator, $\oprho$, the reduced operator, $\oprho_1$, describes the same
probability as the full density operator for all measurements concerning the subsystem-1 .
\footnote{A measurement on subsystem-2 only affects $\oprho_1$ 
if the information on the outcome of the measurement is made available \cite{N-C}.}

The different forms of the partial trace depend essentially on
the Hilbert-Schmidt product \eref{HJprod} of each basis with the identity.
In the case of the position basis, we have
\be
\langle\!\langle I |Q\rangle\!\rangle= \rm tr\>\opI\>\>|q_-\rangle\langle q_+|
=\delta(q_- -q_+),
\ee 
so that
\be
\fl A_1(Q_1)= \int \rmd q_{2-} \rmd q_{2+} \>\>A(Q_1, Q_2\!=\!(q_{2-},q_{2+}))\>\>\delta(q_{2-} -q_{2+})
        =\int \rmd q_2 \> \>A(Q_1,(q_2,q_2)).
\ee
Here, we should recall that,
\be
A(Q_1,(q_2,q_2))=\langle q_{1-},q_2|\>\widehat A\>|\>q_{1+},q_2\rangle,
\ee
in matrix notation.

In the centre representation, we have simply
\be
\langle\!\langle I |\x\rangle\!\rangle= {\rm tr}\>\opI\>\>(2^L\opR_\x)=1,
\ee 
leading to the phase space projection:
\be
A_1(\x_1)=\int \rmd \x_2 \> \>A(\x_1,\x_2).
\ee
In the case of the density operator, the corresponding reduced Wigner function, $W_1(\x_1)$,
is thus obtained from $W(\x)$ in the same way as a marginal
probability distribution is projected out of the full distribution \eref{marginal}.
 
The simplest choice turns out to be the chord representation.
Then, $|I\rangle\!\rangle=\opT_{\vxi=0}$ is an element of the 
operator basis, so that
\be 
\langle\!\langle I|\y\rangle\!\rangle=\delta(\y)=\delta(\vxi).
\ee
Thus in this case, instead of projecting, we obtain the reduced
operator merely by slicing through the chord symbol:
\be
A_1(\vxi_1)=A(\vxi_1,\vxi_2\!=\! 0).
\label{redchord}
\ee
Of course, the reduced operator $\opA_1$ itself is insensitive to
the procedure used to obtain it within the various representations, 
but the ease of calculating the reduction is a special bonus 
of the chord representation. 

It should be recalled that the partial trace is invariant
with respect to unitary transformations performed internaly
within the factor Hilbert space ${\bi H}_2$: $\opU=\opU_2\otimes\opI_1$
(see e.g. \cite{N-C}). In the example where the subsystems are particles
that have separated by a large distance, then these are truly {\bf local transformations}.
In other words, if $\opA'=\opU\opA\opU^{-1}$, then
$\rm tr_2\>\opA'=\rm tr_2\> \opA$. This invariance corresponds semiclassically
to the freedom of performing canonical transformations which
leave invariant the $\x_1$ variables: $(x_1,x_2)\rightarrow(x_1,x'_2)$.
This also implies that only the double phase space corresponding to $x_2$
changes: $(X_1,X_2)\rightarrow(X_1, X'_2)$. If the canonical transformation
is linear in the single phase space, then both the
centres, $\x$, and the chords, $\vxi$, are propagated 
in the arguments of their respective functions by this same transformation. 

Another point that is worth discussing concerns the completeness 
of the operator representations. Notice that the restricted translation operators
\be
|\y_1\rangle\!\rangle_1= \opT'_{\vxi_1}=\opT_{\vxi_1}\otimes \opI_2
\label{restrans}
\ee
are a subset of the translation operators used in the chord basis
for the full Hilbert space, ${\bi H}_1\otimes {\bi H}_2$. It follows that
a representation in terms of the restricted translation operators,
$\opT'_{\vxi_1}$, would not be complete. Likewise, we may define the
restricted unitary reflection operators,
\be
|\x_1\rangle\!\rangle_1= 2^{L_1}\opR'_{\x_1}=\opR_{\x_1}\otimes \opI_2,
\label{resrefl}
\ee
but these do not belong to the centre basis for ${\bi H}_1\otimes{\bi H}_2$.
Even so, we may also define directly the reduced operator $\opA_1$ as
\be
\opA_1=\int \rmd \x_1 \>A_1(\x_1)\>|\x_1\rangle\!\rangle_1,
\label{redop}
\ee
with
\be
A_1(\x_1)={\rm tr}\> \opA \>(2^{L_1}\opR'_ {\x_1}).
\label{redweyl} 
\ee  

Let us now specialize to density operators. 
In the case of the chord function,
we must take care of the normalization, which depends on the
number of degrees of freedom. Hence, the validity of \eref{redchord}
between the chord representation of density operators, 
$\rho(\vxi)$ and $\rho_1(\vxi_1)$, 
implies that the reduced chord function is
\be
\chi_1(\vxi_1)=(2\pi\hbar)^{L_2}\>\chi(\vxi_1,\vxi_2\!=\! 0).
\label{redchord}
\ee
Clearly, $\chi_1(\vxi_1)$ is the Fourier transform of $W_1(\x_1)$.
Since the definition of phase space correlations \eref{pscor} 
is valid for the reduced system, we obtain the reduced correlations
as a projection of the correlations of the entangled pure state:
\begin{eqnarray}
C_1(\vxi_1) & &=
{\rm tr}\> \oprho_1 \>\widehat{T}_{(\vxi_1,0)}\>\oprho_1\>\widehat{T}^\dagger_{(\vxi_1,0)}\nonumber \\
& &=(2\pi \hbar )^{L_1} \int \rm d\veta_1 \; e^{ i {\veta}_1 \wedge {\vxi}_1 / \hbar}
\left| \chi_1({\veta}_1 )\right|^2\nonumber \\
& &=(2\pi \hbar )^{L_1} \int \rm d\veta_1 \rm d\veta_2 \; e^{ i {\veta_1} \wedge {\vxi_1} / \hbar}
\delta(\veta_2) \>\left| \chi(\veta )\right|^2\nonumber \\
& &=\int \frac{\rm d\vxi_2}{(2\pi \hbar )^{L_2}}C(\vxi) \; .
\end{eqnarray}
It should be recalled that the relation between the Wigner function
and the chord function mimics that between a classical probability
distribution and its characteristic function. The definition
of correlations and the classical marginal distributions also goes
through as above. Therefore, the property that the correlation
of the reduced state for a given displacement, $\vxi_1$,
is just the integral over all correlations in the larger space
over displacements that share this component also holds for
classical probability distributions. This relation does not
depend on the full density operator being a pure state.

All the representations
that we have been discussing will factor in the case that 
$\oprho=(|\psi_1\rangle\otimes|\psi_2\rangle)(\langle\psi_2|\otimes\langle\psi_1|)$
is a product pure state. Thus we obtain product Wigner functions, $W(\x)=W_1(\x_1)\>W_2(\x_2)$ 
and product chord functions, $\chi(\vxi)=\chi_1(\vxi_1)\>\chi_2(\vxi_2)$. 
These relations may be interpreted in terms of average values of the basis operators, i.e.
$\langle \opR_\x\rangle=\langle \opR'_{\x_1}\rangle\>\langle \opR'_{\x_2}\rangle$ and
$\langle \opT_\vxi\rangle=\langle \opT'_{\vxi_1}\rangle\>\langle \opT'_{\vxi_2}\rangle$.
Thus, a sufficient criterion for the existence of entanglement would be that either of these  
equalities not hold for some centre, $\x$, or some chord, $\vxi$.  

Curiously, it is not the generation of cross correlations
that is usually taken as a measure of entanglement, but instead
the loss of correlations of the reduced density operator.
Its von Neumann entropy (\ref{vonneumann}) is often referred to as {\bf the entanglement}.
Expanding this to first order, results in the linear entropy of
of a partial trace of the full density operator, 
\be
1-{\rm tr}\>\oprho_1^2=1-C_1(0), 
\ee
recalling (\ref{pscor}), which is the square of the {\bf concurrence},
another widely used entanglement measure (see e. g. \cite{MinBuch}).
This is not an obvious measure of overall entanglement, because we should
obtain the same measure by singling out instead the reduced density operator for subsystem-2.
But, it is a simple consequence of \eref{Finvariance}, 
the invariance of the quantum correlations with respect to Fourier transforms
for a pure state, that
\begin{eqnarray}
C_1(0)
& &=\int \frac{\rmd\vxi_2}{(2\pi\hbar)^{L_2}}C(0,\vxi_2)\nonumber \\
& &= \int\frac{\rmd\vxi_2}{(2\pi\hbar)^{L_2}} \int \frac{\rmd\eta}{(2\pi\hbar)^L} 
\>C(\eta) \>e^{i\eta_2 \wedge \xi_2}\nonumber \\
& &= \int \frac{\rmd\eta_1}{(2\pi\hbar)^{L_1}}C(\eta_1,0)
= C_2(0).
\label{purity}
\end{eqnarray}
Reinterpreted in terms of Wigner functions,
\be
\int \rmd \x_1\> [W_1(\x_1)]^2=\int \rmd \x_2\> [W_2(\x_2)]^2,
\ee
this is another remarkable property of pure quantum states,
for it is highly unusual for the second moment of a pair
of marginal probability distributions to display a similar equality.
Indeed, it is not even generally true for product distributions.

The focus on properties of the reduced density matrix
makes sense when it is recalled that the concept of entanglement
involves separate measurement on each of the components.
The invariance of the partial traces with respect to local transformations
carries over to the above measures of entanglement. Even better,
it has been shown that it is possible to concentrate the entanglement
within a few elements of an ensemble of identical states,
by performing local measurements \cite{BBPS}.

In terms of Husimi functions (\ref{Husimi}), it is natural to describe
entanglement in terms of the Wehrl entropy (\ref{Wherl}) 
for the reduced density operator. 
Another way of describing entanglement is through the Schmidt decomposition
\eref{Schmidt}. The corresponding Wigner and chord functions are then
\be
W(\x)=(\pi\hbar)^{-L}\sum_{i,j}\lambda_i \lambda_j \> 
\langle{\psi^i}_1|\opR'_{\x_1}|{\psi^j}_1\rangle \langle{\psi^i}_2|\opR'_{\x_2}|{\psi^j}_2\rangle
\ee
and
\be
\chi(\vxi)=(2\pi\hbar)^{-L}\sum_{i,j}\lambda_i \lambda_j \> 
\langle{\psi^i}_1|\opT'_{-\vxi_1}|{\psi^j}_1\rangle \langle{\psi^i}_2|\opT'_{-\vxi_2}|{\psi^j}_2\rangle,
\ee
recalling the definitions of the restricted reflection operators \eref{resrefl}
and the restricted translation operators \eref{restrans}.
In both cases the partial trace over subsystem-2 
substitutes the second Dirac bracket by $\delta_{i,j}$, so that
\be
W_1(\x_1)=(\pi\hbar)^{-L_1}\sum_i{\lambda_i}^2 \> 
\langle{\psi^i}_1|\opR'_{\x_1}|{\psi^i}_1\rangle
=\sum_i{\lambda_i}^2 \>W_i(\x_1) 
\ee
and
\be
\chi_1(\vxi_1)=(2\pi\hbar)^{-L_1}\sum_i{\lambda_i}^2 \> 
\langle{\psi^i}_1|\opT'_{-\vxi_1}|{\psi^i}_1\rangle
=\sum_i{\lambda_i}^2 \>\chi_i(\vxi_1).
\ee
Therefore, the reduced density operator is just a mixture 
of the factor states in the Schmidt decomposition for subsystem-1,
with probabilities specified by the square of the Schmidt coefficients.
The concurrence is then given by
\be
1-{\rm tr}\>(\oprho_1)^2= 1 - \sum_i{\lambda_i}^4,
\ee
in terms of the second moment of the weighing factors for the mixed state.
Note that, contrary to the Schmidt number, this is a well defined
entanglement measure for systems with infinite Hilbert spaces,
if the above sum converges.
Clearly, the purity of subsystem-2 involves the same sum over Schmidt
coefficients, in agreement with our previous calculation \eref{purity}.

Consider now the case that a subsystem 
can again be split up into a pair of components. 
If the full original state was entangled, the reduced density operator
is not pure. Hence, it is an average over pure states.
Obviously, this cannot be a product state overall,
but if all of the pure states are products, the mixed state is not
characterized as entangled, rather it is a {\bf separable state}.
The problem with mixed states is that the decomposition into pure states
is not unique, so a state is considered separable if there exists any
decomposition where it is separated. 

Let us now define a {\it classical pure state} as a $\delta$-function in phase space.
Then all pure states, $f(x)=\delta(x)$, in a higher dimensional phase space 
will be product states, because the higher dimensional $\delta$-functions 
factor as $\delta(x)=\delta(x_1)\>\delta(x_2)$. In this sense, the expression
\be
f(x)=\int f(x')\> \delta(x-x') {\rm d}x'
\label{classep}
\ee
can be reinterpreted as a {\it classical separable state}:
Any probability distribution in phase space can be considered
as a linear combination of products of classical pure states.
Thus, we can never consider a classical phase space distribution to be entangled,
no matter how strong the correlations may be between variables pertaining to different
subsystems.

What if a mixed Wigner function for a bipartite state is everywhere positive? 
Can we mimmick the above reasoning to conclude that there is no entanglement? 
In general this is not so,
because the function $\delta(\x)$ does not represent a density operator in the 
Weyl representation. It represents instead the reflection operator, 
which has an infinitely degenerate negative eigenvalue, as discussed in section 5. 
The closest that is possible is the coherent state (\ref{wcoherent}),
which approaches a $\delta$-function as $\hbar\rightarrow 0$,
but imposes an extra smoothing on the Wigner function for any combination of
these pure states. Indeed, a general superposition of coherent states is defined by
a weight function, known as the Glauber-Sudarshan {\bf P-function} 
in quantum optics \cite{Glauber, Sudar, Schleich}.
So, it is the positivity of a P-function that guarantees a separable state,
rather than that of the Wigner function, because each coherent state can be factored. 
 
To close this section, let us now study 
another kind of projection of the Wigner function.
Whereas, by projecting onto a component subspace we
generate a reduced Wigner function, a projection onto
a Lagrangian plane (\ref{qproj}) results in a probability density.
All the coordinates of such a plane correspond to commuting
operators. In the case of a bipartite system, 
we can define this Lagrangian plane by choosing 
some linear combination of the variables for each subsystem, 
$q'_1=\alpha_1 p_1+\beta_1 q_1$ and $q'_2=\alpha_2 p_2+\beta_2 q_2$,
so that each coordinate, $q'j$, pertains to a diferent subsystem.
 
Consider now pairs of either-or measurements on both these variables,
i. e. we can define observables 
$\widehat O_{1a}$, $\widehat O_{1b}$, $\widehat O_{2a}$ and $\widehat O_{2b}$
which take the value $+1$, for $q'_j$ in the interval $ja$, and $-1$ outside.
In terms of projection operators $\widehat P_{ja}$, we have
$\widehat O_{ja}=2\widehat P_{ja}-\>\widehat I$ and
$\widehat O_{1a}\widehat O_{2a}=
4\widehat P_{1a}\widehat P_{2a}- 2\widehat P_{1a}-2\widehat P_{2a}+\widehat I$,
with similar formulae for the other products of commuting operators.
Combining the expectation values for these products in the form of 
the CHSH inequality (\ref{CHSH}),
\begin{eqnarray}
\fl\langle \widehat O_{1a}\widehat O_{2a} \rangle + \langle \widehat O_{1a}\widehat O_{2b} \rangle
+ \langle \widehat O_{1b}\widehat O_{2a} \rangle - \langle \widehat O_{1b}\widehat O_{2b} \rangle
=&4[\langle \widehat P_{1a}\widehat P_{2a} \rangle + \langle \widehat P_{1a}\widehat P_{2b} \rangle
+ \langle \widehat P_{1b}\widehat P_{2a} \rangle - \langle \widehat P_{1b}\widehat P_{2b} \rangle]\nonumber\\
&-4[\langle \widehat P_{1a}\rangle+\langle\widehat P_{2a} \rangle]+2,
\end{eqnarray}
we can now evaluate each expectation value on the right hand side as a definite integral
of the probability density over some region of the $(q'_1, q'_2)$ plane. 
This is a purely classical setup, so that by regrouping,
\be
\fl 2-4[\widehat P_{1a}-\langle \widehat P_{1a}\widehat P_{2b} \rangle]
-4[\widehat P_{2a}-\langle \widehat P_{1b}\widehat P_{2a} \rangle
-\langle \widehat P_{1a}\widehat P_{2a} \rangle]
-4\langle \widehat P_{1b}\widehat P_{2b} \rangle \leq 2,
\ee
we rederive the CHSH inequality,
because the square brackets cannot be negative.

We thus verify that the correlations measured among commuting
pairs of either-or observables of each subsystem lie within strictly classical bounds,
irrespective of any possible entanglement of their combined state.
It makes no difference whether, or not, the Wigner function has negative regions.
The point is that we need only deal with a single positive projection,
which is a true probability distribution. To obtain a violation of the
CHSH inequality, we must choose pairs of observables for each component
which do not commute. The correlation for a given choice of observables
from each pair may still be computed from the probabilities 
in the respective Lagrangian plane, but we must use different planes in each
of the four correlations. Then, if the overall Wigner function that generates
all these densities has negative regions, the CHSH inequality may be violated,
as discussed in the following section.

Apparently, there has not been much effort to relate the intuitively
appealing picture of entanglement as the source of nonclassical correlations
in Bell inequalities to the technical entanglement measures appropriate to
quantum information theory. However, a recent paper by Cirone \cite{Cirone}
bridges this gap for finite dimensional systems. The main point is that
measurements are restricted to projection operators for the factor states
in the Schmidt basis. It is then shown that the same concurrence, which was
introduced in terms of the partial trace, can be expressed as a sum over
conditional probabilities for measurements on either component.

\section{Generating a $classical$ entanglement: The EPR state}

We have seen how symplectic transformations correspond 
exactly to unitary transformations in Hilbert space. Let us now
examine how these can produce entangled states, given that the initial state, $\oprho$,
is a product of states, each represented by its Wigner function, $W_j(\x_j)$,
or its chord function, $\chi_j(\vxi_j)$, so that $W(\x)=W_1(\x_1)W_2(\x_2)$ and
$\chi(\xi)=\chi_1(\xi_1)\chi_2(\xi_2)$.
For the canonical transformation to be linear,
the  classical interaction Hamiltonian, 
$H(x_1,x_2)$ can only be bilinear in the phase space variables.
A convenient choice is
$H=p_1q_2 -p_2q_1$, which may be interpreted as angular momentum,
$L_3$, if the degrees of freedom refer to Cartesian coordinates in a plane.
This Hamiltonian merely rotates both $p$ and $q$ coordinates
in the argument of $W(\x)$ and $\chi(\vxi)$.
Then, after a rotation by $\pi/4$, the density operator becomes $\oprho'$, represented by 
\be
\chi'(\xi)=\chi_1(\frac{\xi_{p_1}+\xi_{p_2}}{\sqrt 2},\frac{\xi_{q_1}+\xi_{q_2}}{\sqrt 2})
\>\>\chi_2(\frac{\xi_{p_1}-\xi_{p_2}}{\sqrt 2},\frac{\xi_{q_1}-\xi_{q_2}}{\sqrt 2}).
\ee
Since the partial trace is specified by (\ref{redchord}), a section of the chord function,
the reduced density for the first component becomes
\be
\chi'_1(\xi_1)=(2\pi\hbar)\; \chi_1(\frac{\xi_{p_1}}{\sqrt 2},\frac{\xi_{q_1}}{\sqrt 2})\>\>
\chi_2(\frac{\xi_{p_1}}{\sqrt 2},\frac{\xi_{q_1}}{\sqrt 2}),
\ee
in the chord representation.

So as to emphasise how {\it classical} an entanglement can be,
let us choose for example initial Gaussian states,
the product of harmonic oscillator ground states, described by
\begin{equation}
 W_j(\x_j) =
 \frac{1}{\pi \hbar}
 \exp \left[
-\frac{\omega_j}{\hbar}    \ \mathbf  q_j^2 -
 \frac{1}{\hbar \omega_j}  \ \mathbf p_j^2      
     \right], 
\label{factors}
\end{equation}
or
\begin{equation}
 \chi_j(\xi_j )= 
\frac{1}{2\pi \hbar}
 \exp \left[-\frac{\omega_j}{\hbar}
 \left(\frac{\xi_{q_j}}{2}\right)^2-\frac{1}{\hbar\omega_j}
 \left(\frac{\xi_{p_j}}{2}\right)^2\right]. 
\end{equation}
Thus, the probability distribution for positions,
\be
f(\mathbf q)= \int \rmd \mathbf p \> W(\x),
\ee
is also a Gaussian with elliptic level curves, 
that are also rotated if $\omega_1\neq\omega_2$.
In this case, the effect of rotation, followed 
by the partial trace, is just a
narrowing of the Gaussians in the chord representation:
\be
\chi'_1(\xi_1) = 
\frac{1}{2\pi \hbar} 
 \exp \left[  -\frac{\omega_1 + \omega_2}{2\hbar}
 \left( \frac{\xi_{q_1}}{2}\right)^2
-\frac{1}{2\hbar}\left({1\over{\omega_1}}+{1\over{\omega_2}}\right)
 \left(\frac{\xi_{p_1}}{2}\right)^2 \right]. 
\ee
Notice that normalization is maintained, because we still
have $\chi'_{1}(\xi_1)=(2\pi\hbar)^{-1}$ at the chord origin, but now the widths of the
position Gaussian and of the momentum Gaussian 
are obtained through different kinds of average. 
The overall narrowing indicates that this is no longer a pure state. 

The Wigner function presents a more intuitive picture of a mixed state.
Taking the Fourier transform:
\be
W'_1(\x_1)= \frac{1}{\pi \Delta}\;
 \exp \left[-\frac{2\omega_1\omega_2 \mathbf  q_1^2} {\hbar (\omega_1 + \omega_2)} -
 \frac{2\mathbf p_1^2} {\hbar (\omega_1 + \omega_2)}     
     \right]  \; .
\label{redgauss}
\ee
This still integrates to one, as demanded by normalization, but
the Gaussian is now broader, with the uncertainty 
$\Delta=(\omega_1 + \omega_2)/2\sqrt{\omega_1\omega_2}>\hbar$,  
if $\omega_1\neq\omega_2$. Therefore, this is not a pure state.
The way that this example relates entanglement to 
initial states and evolution, which may both be considered {\it classical},   
is even more extreme than those discussed
in \cite{AngFur}, which relie on projections of the
Husimi function, in the approximate role of phase space probability density.

Another confirmation that this is not a pure state is that
\be
\rm tr\>(\oprho'_1)^2=2\pi\hbar \int \rmd\x_1 \>[W'_1(\x_1)]^2 = 
2\pi\hbar \int \rmd\vxi_1 \>|\chi'_1(\vxi_1)|^2 <1,
\ee
and yet it might seem that this is just a freak result.
After all, our state has remained a smooth {\it classical-like}
Gaussian throughout. There are none of the quantum oscillations
which are supposed to be the fingerprint of nonclassicallity:
For a start, nothing would prevent us from identifying the original Wigner
function with a classical probability distribution. 
We then perform a simple rotation with perfect
classical correspondence and obtain a new Gaussian,
which pretends to be a quintessentially quantum entangled state!
Have we been led astray? 

Let us go back to the full Wigner function, resulting from the choice (\ref{factors})
of Gaussians for the initial product state. After the $\pi/4$ rotation, this is just 
\begin{equation}
\label{prod}
 \fl W'(\x_) =\left(\frac{1}{\pi \hbar}\right)^2
 \exp \left[-\frac{\omega_1}{2\hbar}    \ \mathbf  (q_1+q_2)^2 -
 \frac{1}{2\hbar \omega_1}  \ \mathbf (p_1+p_2)^2 \right] 
 \exp \left[-\frac{\omega_2}{2\hbar}    \ \mathbf  (q_1-q_2)^2 -
 \frac{1}{2\hbar \omega_2}  \ \mathbf (p_1-p_2)^2 \right]. 
\label{fullW'}
\end{equation}
In the extreme limit where $\omega_1\rightarrow 0$ and $\omega_2\rightarrow \infty$,
we would obtain a normalized version of
\be
W'(x)=\delta(q_1-q_2)\>\delta(p_1+p_2),
\ee
which is just the Wigner function derived by Bell \cite{BellWig}
for the original EPR wave function \cite{EPR}, 
namely $\langle q|\psi\rangle=\delta(q_1-q_2)$.
It so happens that the rotation that transformed the coordinates
of our initial state, i. e. the ground state of an anisotropic plane harmonic oscilator,
is essentially the same as the transformation from the individual coordinates 
for a pair of particles into a centre of mass, together
with a relative {\it internal} coordinate. 
(These transformations differ only by local unitary transformations.) 
The EPR state is a $\delta$-function
both in the relative position and in the total momentum,
which is the conjugate variable to the centre of mass.

Thus, the entanglement verified in our initial example
implies that the centre of mass is likewise entangled with the relative coordinate
in the EPR state. Perhaps, it is then even more surprising 
that the example that was picked is in some sense {\it classical},
if we consider that the discussion of the nonlocal and hence nonclassical
nature of quantum mechanics started off with the historic EPR paper \cite{EPR}.
The fact that the full Wigner function is positive, not only allows us
to interpret it as a classical probability distribution, but it also ensures that
there is a wide range of measurements that can be performed on
either component which may be considered as classical and hence local.
We already found in the previous section that
any measurement of pairs of either-or variables,
$\widehat O_{1a}$, $\widehat O_{1b}$, $\widehat O_{2a}$ and $\widehat O_{2b}$
which take the value $+1$, for general phase space coordinates, 
$q'_j$, in the interval $ja$, and $-1$ outside, have correlations that
satisfy the CHSH inequality, even if the Wigner function has negative regions.
That was the case where the quantum observables which are measured commute. 
The statement for positive Wigner functions, due to Bell \cite{BellWig},
is stronger: The inequality is then satisfied even if we choose different
variables for each measurement, $q'_{1a}\neq q'_{1b}$ and  $q'_{2a}\neq q'_{2b}$,
corresponding to different Lagrangian planes in phase space and, hence, 
quantum operators that do not commute. The argument is essentially
the same as in the last section, except that now we can obtain all the expectation values 
from the full Wigner function, acting as a global probability distribution,
instead of dealing with different probability distributions, each restricted
to the Lagrangian plane specific to a given pair of variables.
 
Let us now reexamine our classically entangled states from the point of view
of the reduced reflection operators, $\opR'_{\x j}$, defined as (\ref{redweyl}), 
that act on each component and, in particular, the parity operators, $\opR'_{0j}$.
Such observables do not correspond to 
smooth phase space fuctions in classical mechanics,
indeed, the Weyl representation of these operators (\ref{weylrefl}) is singular.
Nonetheless, parity, or reflection measurements can also be carried out on classical
waves, as discussed in section 2, and the question now concerns
the possible correlations between measurements for different reflections
carried out on both components.
The fact that the full Wigner function (\ref{fullW'}) is symmetric with respect to the
origin implies that the density operator commutes with  the full reflection operator, $\opR_0$.
However, $W'_1(0)<\pi\hbar$, so it does not have pure parity, 
i. e. $\oprho'_1$ does not commute with $\opR'_{01}$.
Hence, according to the discussion in section 6,
there is a finite probability to obtain negative (odd) parity,
if such a measurement is performed on subsystem-1.

Perhaps this would not be so obvious a priori:
The original state, represented by $W_0(\x)$,
is a pure state with pure positive (even) parity
and this is also a property of the rotated state.
This  property can be verified directly, or it may be noticed that
the driving Hamiltonian commutes with $\opR_0$,
so that $H(\x)=H(R_0(\x))$. But now we find that
a measurement of the parity of subsystem-1 has 
a finite probability to be negative. How is that?

Notice that the same also hods for subsystem-2:
The derivation of the reduced density operator, $\oprho'_2$,
represented by $W'_2(\x_2)$ and $\chi'_2(\vxi_2)$,
goes through exactly as above. Therefore there is also
a finite probability of measuring negative parity in subsystem-2.
As was shown in section 6, the fact that, 
in both cases, the Wigner function is symmetric about the origin
implies that all the pure states, into which the mixed reduced 
density operator can be decomposed, must have pure parity, but they are not
all even. For this reason, the Wigner function (\ref{redgauss}) had to be obtained as
a Fourier transform of the chord function; not a mere rescaling.

The crucial point is that the rotated state, $\oprho'$, 
does not commute with  either of the restricted reflections defined by \eref{resrefl},
i.e. $\opR'_{0 1}$ or $\opR'_{0 2}$, even though it commutes with their product:
$\opR_0=\opR'_{0 1}\opR'_{0 2}=\opR'_{0 2}\opR'_{0 1}$.
It should be recalled that the reduced Wigner functions are entirely determined
by \eref{redop} and \eref{redweyl} in terms of the restricted reflections.
Thus, to understand the results of measurements of either $\opR'_{0 j}$,
we need a common basis for all these operators. This is just the
product of an even-odd basis for subsystem-1 and subsystem-2,
for which we obtain the table:
\be
even\otimes even\rightarrow even
\ee
\be
even\otimes odd\rightarrow odd
\ee
\be
odd\otimes even\rightarrow odd
\ee
\be
odd\otimes odd\rightarrow even
\ee
Since $\oprho'$ is even, it must be a superposition of the subset
of basis states: $even\otimes even$, or $odd\otimes odd$.
Furthermore, we now find that the evolved state has a full parity correlation:
If the measurement of $\opR'_{0 1}$ specifies even parity, 
then this must be the outcome of a measurement on $\opR'_{0 2}$.
Conversely, if one of the subsystems has odd parity, then we know
this to be the parity of the other subsystem.

An initial product state 
of an even Schr\"odinger cat state with a coherent state,
which is rotated by $\pi/4$, is also susceptible to the foregoing analysis.  
However, an odd symmetry Schr\"odinger cat
would have perfectly anticorrelated odd-even, or even-odd subsystems.
In the case of the rotated cat the evidence for entanglement is much more
obvious. The pair of Gaussians is not centred on either of the planes
in the chord phase space pertaining to the pair of subsystems.
The partial trace that generates the reduced chord functions is
a section of the full chord function, so that it does not capture these
local maxima. Therefore, there is a deficit of phase space correlations in the
reduced density operators.

Returning to the original rotated squeezed state,
or, equivalently, the original EPR state,  
we must conclude that this is truly quantum 
and correctly described as entangled, i. e. just as
nonclassical as the spin states in the Bohm version of EPR \cite{Bohm} 
that are commonly used to exemplify entanglement.
The secret lies in choosing the property to be measured:
A position measurement on one of the subsystems would not
distinguish between this pure quantum state and a classical
distribution. However, a measurement of reflection eigenvalues
evokes a {\it spin-like duality} of this apparently classical state.

The violation of the CHSH inequality for reflection measurements 
of the smoothed EPR state completes the evidence of its nonclassicality.
Banaszek and Wodkiewicz \cite{BK1} first pointed out that 
the full pure state Wigner function of a bipartite state 
is proportional to the correlation for relection measurements on each subsystem:
$\langle \opR'_{\x 1}\opR'_{\x 2}\rangle=(\pi\hbar)^2W(\x)$. 
This leads to a violation of the CHSH inequality for reflection measurements of the EPR state.
They also proposed a realistic experiment for this in quantum optics \cite{BK2}.
We have already verified the complete correlation for parity
measurements about the origin, which is in agreement with the maximal value 
that the full Wigner function (\ref{fullW'}) attains there.
Its decay for large $\x_1$ or $\x_2$ signifies that
$\langle \opR'_{\x 1}\opR'_{\x 2}\rangle\rightarrow 0$,
so that
\be
C_{CHSH}=\langle \opR'_{0 1}\opR'_{0 2}\rangle+\langle \opR'_{0 1}\opR'_{\x 2}\rangle
+\langle \opR'_{\x 1}\opR'_{0 2}\rangle-\langle \opR'_{\x 1}\opR'_{\x 2}\rangle
\ee
sinks from, 2, its maximal classical value at the origin to the limiting value 1.
However, the origin is not the maximum of $C_{CHSH}(\x_1, \x_2)$,
because the lowest order expansion of
\be
\fl({\pi \hbar})^2 W'(\x)=1
-\frac{\omega_1}{2\hbar}\ \mathbf  (q_1+q_2)^2 - \frac{1}{2\hbar \omega_1}\ \mathbf (p_1+p_2)^2
-\frac{\omega_2}{2\hbar}\ \mathbf  (q_1-q_2)^2 - \frac{1}{2\hbar \omega_2}\ \mathbf (p_1-p_2)^2+...
\ee
leads to
\be
C_{CHSH}(\x_1, \x_2)= 2+(\omega_1-\omega_2)q_1 q_2
+\left({1\over{\omega_1}}-{1\over{\omega_2}}\right)p_1 p_2+...
\ee
Hence, the origin is a saddle point of $C_{CHSH}(\x_1, \x_2)$, which {\it increases} from its
maximal classical value along the directions $q_2=-q_1$ and $p_2=p_1$,
if one chooses the EPR conditions, $\omega_2>>\omega_1$, i.e if reflections
are chosen in the directions where the Wigner function decays rapidly.

So we find that nonlocal correlations between two subsystems can arise
even if the Wigner function for the full system is everywhere non-negative.
It would thus appear that there is no relation between fringes
in the Wigner function, where it attains negative values, and entanglement.
The former project as interference fringes for possible measurements,
but this is quite a different kind of {\it nonclassicality} than the 
delicate nonclassical correlations resulting from entanglement.
But even here, one must be wary! If the measurements on the different
components concern mechanical observables, natural for
classical particles, then there is at least one case where negativity
of the Wigner function has been shown to produce nonclassical correlations.
Indeed, Bell \cite{BellWig} constructed an example where the CHSH inequality
is violated for measurements on pairs of different variables, $q'j=qj+t_j p_j$
and $q'j=qj+\tau_j p_j$. \footnote{Even though Bell reffers to the transformation parameters
as {\it times}, these should be understood as specifications of the variables
and hence of the planes onto which the Wigner function is projected.} The
state for which this is proved is a variation of our rotated state,
where one of the factor Gaussians is substituted by the second
excited state of the harmonic oscillator.

It should always be remembered that entanglement is not
an intrinsic property, but only acquires its meaning within a specified basis,
{\it the computational basis}, or the basis where measurements are made. 
In this respect, it resembles semiclassical caustics, 
which depend on our choice of representation.
If the physical realization of the foregoing example 
were the ground state of a 2-D harmonic oscillator, then the rotation, which was
found to produce entanglement, could be dismissed as 
merely an inconvenient coordinate transformation:
Unless all measurements 
were to be restricted to the original coordinate axes,
it would not be relevant, though true, to say that the rotated system became
entangled, while the original system was a mere product.
In contrast, for the alternative physical interpretation of one of the new
coordinates as the centre of mass for a pair
of particles, its entanglement with the internal coordinate can be important.

\section{Entanglement and decoherence}

The process of decoherence also results from the interaction
of a pair of systems; the (small) {\it open system}
and a (large) system, which we call the {\it environment}.
In contrast to the previous example, the component over which we trace,
so as to obtain the reduced density operator, 
is on a scale which defies anything but a statistical description. 
The ususal picture is that the environment lies {\it somewhere outside},
but it may just as well consist of the internal degrees of freedom
for the centre of mass (CM) of a large system of particles.
Exchanges between the large scale motion and the internal
variables lead to macroscopic energy dissipation as well as decoherence 
of the quantum state for the CM.

Let us consider the simplest possible example of 
the decoherence of the CM, because of its entanglement
with internal variables. The CM for a system of $L$ identical particles, 
assumed to be distinguishable is $Q=(q_1+...+q_L)/L$.
The conjugate variable to $Q$ is the total momentum,
$P=p_1+... +p_L$. 
Let us further imagine that 
they are each in the same single particle state, $\oprho$, and that
these are independent, i. e. both the Wigner and the chord function
are products over those of the individual states. This may seem too 
restrictive, because we should allow for different values of each
average position $\langle q_j\rangle$, but we can redefine this
as the origin for each $j$, so that we then measure Q from  $\langle Q\rangle$.
 
In the case of $L=2$, $X=(P,Q)$ is obtained from the rotated coordinate
in the previous section by a mere canonical rescaling of $2^{\pm 1/2}$.
It has been repeatedly emphasised that all such symplectic transformations
on the argument of the Wigner or the chord function correspond exactly to
unitary quantum transformations. So let us now reverse this transformation
in the case of general $L$: We define $Q'=L^{1/2} Q$ and  $P'=L^{-1/2} P$.
Then, if the individual Wigner functions, $W(\x_j)$, were classical probability distributions,
the {\bf Central Limit Theorem} would imply that the distribution for $X'=(P',Q')$ converges to
\be
W_L(\bf X')\rightarrow [\pi\Delta_{\bf K}]^{-1} \exp[-{\bf X'}{\bf K}^{-1}{\bf X'}/2],
\ee
as $L\rightarrow\infty$, where we recall the  definition of the 
Schr\"odinger covariance matrix, $\bf K$, in (\ref{Kov}) and its determinant ${\Delta_{\bf K}}^2$. 
It is remarkable that positivity is not a necessary ingredient for the
proof of the Central Limit Theorem: In the case of identical 
square-integrable pure state Wigner functions, 
it is shown by Tegmark and Shapiro \cite{TegShap}, 
that convergence onto a Gaussian again results. 
If the state for the individual particles is not 
represented by a pure state Gaussian, then the moments
for this state will be such that $\Delta_{\bf K}>\hbar$.
Therefore, the centre of mass Wigner function, $W_L(\bf X)$,
is a broader Gaussian than is permissible for a pure state 
and hence it must be a mixture. So, the CM of 
independent particles with identical Wigner functions is
generally entangled with the internal phase space variables
(which it has not been necessary to describe explicitly).
Curiously, the potential entanglement resulting from
the Central Limit Theorem was overlooked in \cite{TegShap}.

How does this entanglement with the internal coordinates evolve in time?
It is easy to verify that free motion, generated by the Hamiltonian $H={p_1}^2+...+{p_L}^2$,
will not alter $\Delta_{\bf K}$. Thus the entanglement of the centre of mass with
the environment is invariant in this simple case. Let us supose instead
that, though the particles do not interact, there is an external nonlinear field.
Furthermore, the particles are sufficiently separated and the field 
is smooth enough, so that it is legitimate to linearize the field locally 
around each $\langle q_j\rangle$.
Then the Wigner function for each particle will evolve classically in different ways. 
The restrictive form of the Central Limit Theorem in \cite{TegShap} cannot be applied
in this case, but one can readily adapt Levy's proof \cite{Levy} to allow for
different Wigner functions, as long as the moments are finite 
and their average values congerge \cite{AlmMagNem}.

The averages of the moments resulting from the different evolutions of  
many Wigner function lead to a progressive loss of purity for the CM.
Just as in (\ref{redgauss}) for the simple example of the last section, 
the uncertainty, $\Delta_{\bf K}$, increases.
On top of that, the Central Limit Theorem supplies the statistical ingredient
for the decoherence process.  
It might appear strange to obtain decoherence even for a system of noninteracting
particles, but it should be recalled that the CM momentum $P$, or $P'$, appears
linearly in each of the terms, ${p_j}^2$, in the kinetic energy, 
which accounts for the coupling to the internal momenta.

So as to make contact with the theory of Markovian open systems,
we can now reinterpret this evolution of the reduced density matrix
as a convolution of the original (Gaussian) Wigner function for the CM
with a broadening Gaussian. For its Fourier transform, the chord function, 
this evolution is merely the product of an initial Gaussian with another
Gaussian that narrows in time. \footnote{It must be recalled that the average CM evolution,
$\langle X(t) \rangle$ has been hidden by a time dependent coordinate transformation.} 
This is exactly the result for {\it quantum Markovian evolution} of an open system, 
in the case of quadratic internal Hamiltonian 
and linear coupling to the environment \cite{BroAlm04}.

The deduction of the {\bf canonical Lindblad equation} (see e.g. \cite{Giulini}),
\be
\frac{\der\widehat{\rho}}{\der t}=-\frac{i}{\hbar}
\bigl[\widehat{H}, \widehat{\rho}\bigr]
-{1\over{2\hbar}}\sum_j ~
\bigl[\widehat{L}_j, \bigl[\widehat{L}_j, \widehat{\rho}\bigr]\bigr],
\label{def-master}
\ee
that governs the evolution of the density operator in the quantum Markovian theory
does not proceed by tracing out a larger system. 
All the same, the mere fact 
that the evolution is entirely determined by a differential
equation precludes any delayed participation of previous motion.
The {\bf Lindblad operators}, $\widehat{L}_j$, account for the nonunitarity
of the evolution, that is, they take the part of the coupling to the environment.
The Markovian approximation can in principle include arbitrary (non-quadratic)
internal Hamiltonians for the system. 
 
The derivation of the Markovian approximation in the context of quantum optics 
(the damped harmonic oscillator) 
was carried out originally by Agarwal \cite{Agarwal}, 
but this is all in the language of complex phase space. 
The exact solution of (\ref{def-master}) in \cite{Agarwal}
and that of Diosy and Kiefer \cite{DioKief}, for the free {\it open} particle,
are special cases of of the general result in \cite{BroAlm04}:
The chord representation of (\ref{def-master}) is particularly simple if
the Lindblad operators, are linear functions of positions and momenta, 
$\widehat{L}_j=l_j\cdot\widehat x$,
and if the Hamiltonian is quadratic \cite{BroAlm04}:
\be
\frac{\der \chf}{\der t}(\vxi,t) = \left\{ H(\vxi), \chf(\vxi,t) \right\} - 
\frac{1}{2\hbar} \sum_j \left( \vl_j\cdot\vxi \right)^2 ~ \chf(\vxi,t).
\ee 
Here, the first term is the classical Poisson bracket.
The exact solution of this equation factors into the unitary evolution
of the chord function, undistrubed by the Lindblad operators,
and a narrowing Gaussian factor. In the Wigner representation
this becomes a Gaussian smudging of the unitarily evolving Wigner function.
It is remarkable that the Wigner function becomes positive 
after a time that depends only on the parameters of the Lindblad equation, 
regardless of the initial pure state \cite{DioKief,BroAlm04}.

In our simple example of the evolution of the CM, the Lindblad
operator for its one-dimensional motion, should be chosen as
the total momentum $\widehat P$, because this is the variable that 
couples to the internal motion, 
which is hidden within the Markovian approximation.
Even though the Central Limit Theorem supplied the Gaussian factor
of the evolving chord function, the overall Gaussian form for the evolving CM
does not reflect the richness of other possibilities 
for Markovian evolution. However, by considering the entanglement
of a {\it small system} with the CM of a {\it large system}
and following the treatment of the example in the preceding section,
we obtain qualitatively the general Markovian picture.
This allows an interpretation of the Gaussian smoothing
as originating in the multiple small contributions contemplated
in the Central Limit Theorem.                        

The standard way of going beyond the Markovian 
approximation, so as to include memory
kept by the environment of the previous motion of the system, 
is to use the Feynman-Vernon functional \cite{FeyVer},
in the manner exploited by Caldeira and Leggett \cite{CaldLeg}.

\section{A semiclassical picture of entanglement }

A full semiclassical theory of entanglement is still a program for the future,
fascinating but difficult. However, several of the main elements 
are sketched in this concluding section.

For a start, one should note that it is feasible to fit semiclassical
torus states with Gaussian coherent states placed along the classical torus
in a very satisfactory way \cite{Kenfack}. The number of Gaussians required increases with
a fractional power of $\hbar^{-1}$. The important qualitative feature is that
the interference fringes of the Wigner function, near the midpoint 
of the pair of Gaussians composing a Schr\"odinger cat,
have the same wavevector as the similar fringes at the
centre of a geometrical chord of the classical torus. Therefore, in both cases
we can describe very fine interference fringes related to long chords.
It also follows that our preliminary study of entanglement and decoherence
of cat states is not at all irrelevant for understanding the evolution 
of product semiclassical states. Refinements of the fitting procedure
allow even the description of the diffraction effects near caustics \cite{newfit}.

Before analysing product states and their partial trace, 
recall that dyadic operators, $| \psi \rangle \langle  \phi |$,
live in a kind of squared Hilbert space, which
corresponds to a double phase space. 
These operators were shown in section 5
to correspond to a product Lagrangian surface in double phase space,
$\tau _\psi \otimes \tau_\phi$, if each of these states corresponds
to a Lagrangian surface on its own right. Thus, the projection operator,
or pure state density operator, $\widehat{\rho}=| \psi \rangle \langle  \psi |$,  
is just a particular case of this general rule. If the state, $| \psi \rangle$,  
corresponds to a Bohr-Sommerfeld quantized torus of $L$ dimensions in a $2L$-D
phase space, then the  full density operator must correspond to a  
$2L$-D product torus in $4L$-D double phase space. This is in exact analogy
to the way that a product torus describing the state for several particles (\ref{scproduct})
is obtained from lower dimensional tori. Recalling that we can describe
double phase space in terms of the centre coordinates, $\x$, and the conjugate variables,
$\y =\J\vxi$ (\ref{centrechord}), the semiclassical Wigner function, $W(\x)$, 
is then a superposition of complex exponentials, such that each phase
is obtained by integrating $\y(\x)$ along one of the different branches of the torus. 
Even though this approximation breaks down along
caustics, the latter provide ready indication of regions 
where the Wigner function has a large intensity. 

The problem is then to relate the semiclassical Wigner function, 
defined on the centre plane, to classical structures 
that are also portrayed in this same single phase space.
Let us consider first the semiclassical Wigner and chord functions 
in the simplest case where $L=1$.
The Fock states (\ref{Fockst}) are good examples
of semiclassical torus states when the quantum number $n$ is large.
Introducing the asymptotic expression for Laguerre polynomials,
\be
\lim_{n \to \infty} L_n\left(\frac{z^2}{2n}\right) = 
J_0\left(\sqrt{2}z\right) \;, 
\ee
together with the large argument expansion,
\begin{equation}
J_0 (y) \approx 
\frac{2}{\sqrt{\pi y}} \,  \cos \left(y-\frac{\pi}{4}\right) \; ,
\end{equation}
brings the Wigner function (\ref{Fock}) for these states into a semiclassical form.
To understand this, we must investigate the geometry of the double torus
from the point of view of the simpler quantized curve, 
which is just a circle in this case.

Every point on the double torus represents a pair of points
on the quantized curve and vice versa. 
A given pair of points on the quantized curve, $x_\pm$,
defines a geometric chord: $\vxi=x_+ -x_-$. 
Hence, $\y=\J\vxi$ is the chord coordinate on the double torus,
which has the centre coordinate, $\x=(x_+ + x_-)/2$. Obviously,
the exchange of $x_+$ with $x_-$ produces a new chord of the quantized curve
with the same centre, $\x$. Viewed in double phase space, there must always
be pairs of chords of the double torus projecting onto each centre, $\x$.
The symmetry of this surface with respect to the identity plane, $\y=0$,
leads to complex conjugate phase contributions, in line with the above cosine
for the Fock state. Actually this is a general feature: 
Because the Wigner function is real, the chord pairs will always produce
semiclassical contributions adding up to cosines.

\begin{figure}
\includegraphics[width=15cm]{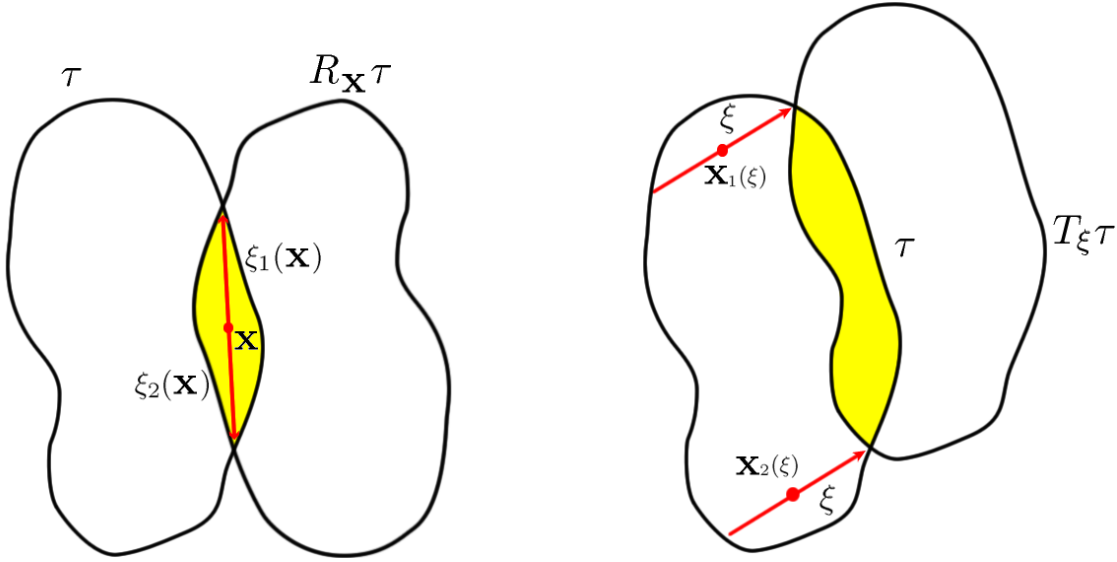}
\caption{(a) The semiclassical Wigner function is constructed by reflecting
the torus $\tau$ around the point $\x$: The chords $\vxi(\x)$ are defined by the intersections
with $R_\x \tau$. (b) The semiclassical chord function is constructed by translating $\tau$
by the vector $\vxi$: The centres $\x(\vxi)$ lie halfway back along $\vxi$ from each intersection
of $\tau$ with $T_\vxi \tau$.}
\label{desenho2}
\end{figure}

To obtain the phase of the cosine contribution to the semiclassical Wigner function
for each pair of chords, the best course is to 
use a result which was put in its most general form by Littlejohn \cite{Littlejohn95}.
This concerns the general overlap, $\langle\psi|\phi\rangle$, of quantum states
associated semiclassically to curves $\tau_\psi$ and $\tau_\phi$:
The semiclassical contributions 
arise from the intersections of these classical curves
and the phase difference between a pair of contributions
is just the area sandwitched between the corresponding pair of intersections,
divided by Planck's constant, as shown in Fig.\ref{desenho2}.

We can immediately apply this principle to the Wigner and chord functions
for pure states, by recalling that
$W_\psi(\x)=(\pi\hbar)^{-1}\langle\psi|(\opR_\x|\psi\rangle)$ and
$\chi_\psi(\vxi)=(2\pi\hbar)^{-1}\langle\psi|(\opT_\vxi|\psi\rangle)$.
The semiclassical state $\opR_\x|\psi\rangle$ is merely the state constructed
from the reflected curve, $R_\x(\tau_\psi)$, whereas $\opT_\vxi|\psi\rangle$
corresponds to the translated curve $T_\vxi(\tau_\psi)$.
Therefore, in the case of the Wigner function, we obtain the phase of 
the cosine as half of the area sandwitched between the torus and its reflection
at the centre $\x$ \cite{AlmHan82}, which coincides with 
the area between the torus and the chord \cite{BerShub,Ber77} (see also. \cite{BerHouches}). 
Furthermore, this construction supplies, at a glance,
the tips of all the chords centred on $\x$ as the intersections between both curves,
as seen in Fig.~\ref{desenho2}(a). 
Note that, once the curve has been reflected around the origin, 
we need only translate $R_0( \tau_\psi)$ to obtain the reflections around all other centres, 
because of the group property,
$R_\x=T_\x \circ R_0 \circ T_{-\x}$.  

The same geometrical method can be used to study the structure
of centres for a pre-specified chord, $\vxi$, on the curve, $\tau_\psi$.
Each intersection of $\tau_\psi$ with the translated curve,  $T_\vxi ( \tau_\psi)$,
reveals one of the tips where the chord is to be placed and hence
the centre of the chord, as shown in Fig.~\ref{desenho2}(b). In the case of open curves, the chord function
may actually be simpler than the Wigner function, 
because it is not necessary to have interference, 
as in the case of the parabola, for which there is only one intersection.
In the case that $\tau_\psi$
has a centre of symmetry, as in the example of the Fock state,
we thus find that the simple relation (\ref{symwig})
between Wigner and chord functions is respected by the semiclassical approximation.

Viewed in single phase space, the caustics of the Wigner function arise
from coalescing torus chords, as their centre, $\x$, is moved. This occurs at the 
tangencies of $R_\x( \tau_\psi)$ with  the fixed curve, $\tau_\psi$. 
Similarly, the caustics of the chord function are the loci of $\vxi$ such that
$T_\vxi ( \tau_\psi)$ is tangent to $\tau_\psi$.
On the other hand, in double phase space, the Wigner caustics
for a torus state are viewed as projection singularities 
of the double torus, $\tau_\psi \otimes \tau_\psi$, which
lies {\it above} the area inside $\tau_\psi$.
The general geometric constructions underlying the semiclassical
Wigner and chord functions are readily extended to phase space representations
of dyadic operators, $| \psi \rangle \langle  \phi |$, corresponding
to double tori, $\tau_\psi \otimes \tau_\phi$. Their Weyl representation
is known as {\bf cross-Wigner functions} or {\bf Moyal brackets} \cite{Moyal},
whose semiclassical form is presented in \cite{Alm84}. 

There are many fascinating features of caustics in the phase space representations
of pure states that have been studied and many more that must still be analyzed.
For instance, the build up in the
centre of the Wigner function for the Fock state is a caustic. 
Its semiclassical origin is the degeneracy 
of a continuum of chords conjugate to the same symmetry centre.
However, this is a nongeneric feature of reflection-symmetric states.
If the symmetry is broken, this supercaustic unfolds into a cusped triangle,
first described by Berry \cite{Ber77}. The unfolding of higher dimensional
caustics for rotated product tori, studied in \cite{AlmHan82} 
were also examined for the Wigner function. It turns out that
the double fold surfaces of the Wigner caustic that meet along the
torus do not unfold in the manner portrayed in Fig.\ref{corner}, because of 
a symmetry constraint.

\begin{figure}
\includegraphics[width=15cm]{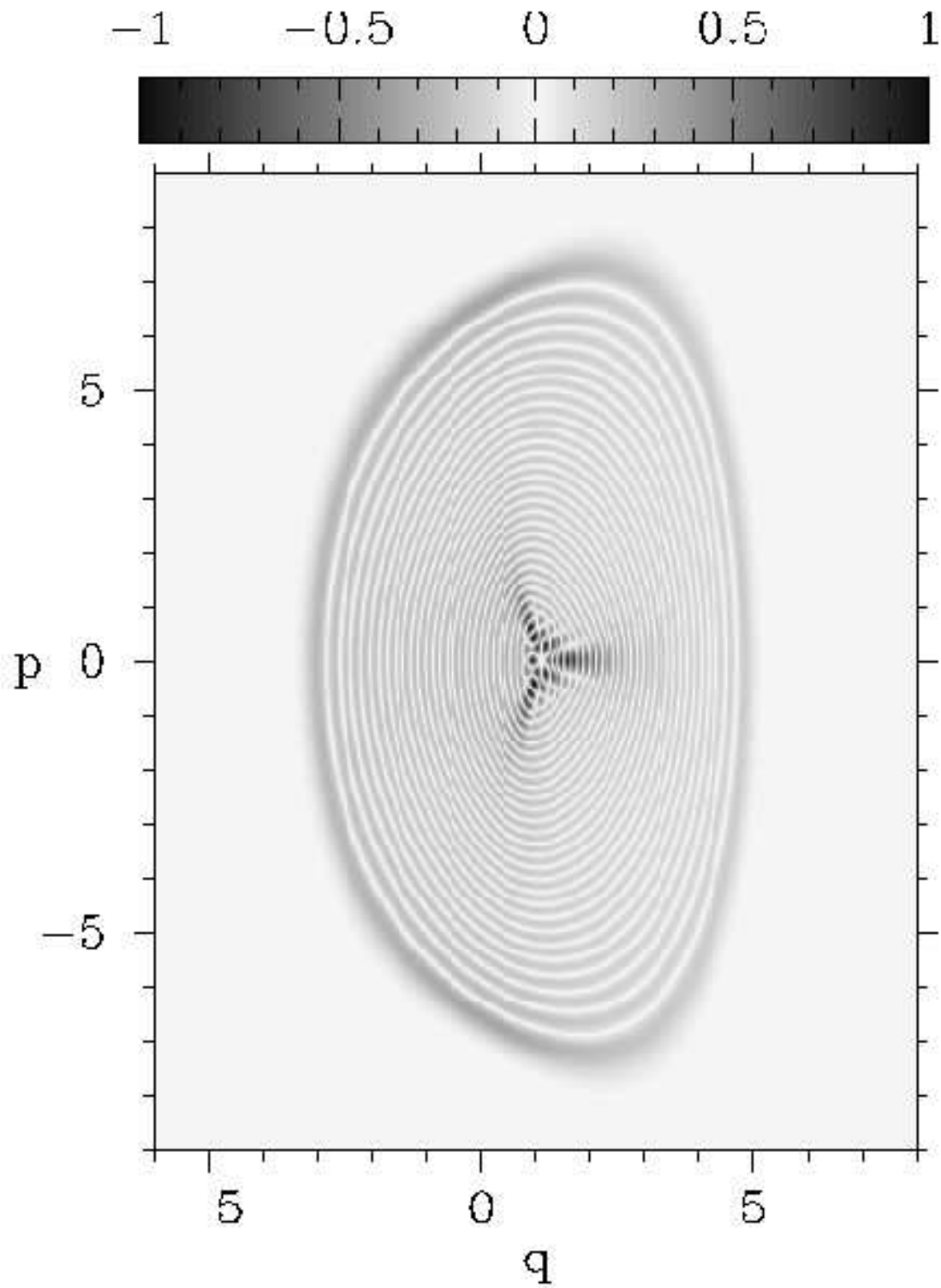}
\caption{Typical fringes for a semiclassical Wigner function.
The classical curve follows closely the border of the fringes. 
The {\it triangular structure} of interfering fringes 
near the centre results from the umfolding 
of the maximum of the Fock states due to symmetry breaking.}
\label{franjas}
\end{figure}

The limit of small chords is specially relevant for semiclassical theory. 
For the Wigner function, it singles out the classical torus itself as the Wigner caustic.
The uniform approximation for the Wigner function throughout this region
is presented in \cite{Ber77}, for the case of a curve 
and in \cite{Alm83}, for a two dimensional torus. 
A pair of sheets of the double torus are joined 
on the identity plane along this curve, or torus. 
The large amplitude of the Wigner function oscillations near
the quantized curve is due to this caustic.  
The corresponding caustic of the chord function
collapses onto the origin, whatever the geometry of the classical region.
The neighbourhood of this highly nongeneric
chord-caustic is discussed in \cite{AlmVal04}.
Once again, we find that all regions where $C(\vxi)=|\chi(\vxi)|^2$ is large,
outside of an $\hbar^{L/2}$-neighbourhood of the origin, 
point to phase space correlations
that are truly quantum in nature. 

So far, we have only discussed static properties of the density operator.
In turning to dynamics, a preliminary point is that 
we should distinguish the {\bf Weyl propagator}, $U(\x)$,
that is, the Weyl representation of the evolution operator, from the propagator
for Wigner functions. The former is unitary and is hence supported
by its own {\it static} Lagrangian surface in double phase space, 
as discussed in section 5, so that its semiclassical description 
is similar to that of the Wigner function itself. This can be
deduced from a path integral in single phase space \cite{Alm92, Report}. 

So as to treat the unitary evolution of the density operator for a pure bound state,
$| \psi \rangle \langle  \psi |$, with $L$ degrees of freedom,
we need to consider the corresponding classical evolution 
of a $2L$-D Lagrangian torus, $\tau_\psi\otimes\tau_\psi$.
Initially, this is separable within both single phase spaces,
even though the product is not factored in 
the centre$\times$chord coordinates. The classical motion 
must propagate the tips of each chord, $x_-$ and $x_+$, in the same way.
Taking account of the change of sign, $p_- \rightarrow -p_-$,
in the definition of double phase space, we find that
the double phase space Hamiltonian must be 
\be 
 I\!\!H(X)=H(x_+)-H(x_-)=H(\x-\J \y/2)-H(\x+\J \y/2).
\ee

This classical Hamiltonian can be verified to preserve the product form of
the geometric structures in each of the phase spaces $x_\pm$,
but it will not preserve initial products within each of these
in the general case that the single Hamiltonian $H(x)$ has
coupling terms between different degrees of freedom.
It propagates Lagrangian surfaces in double phase space
that correspond either to density operators 
(according to the Liouville-von Neumann equation), 
or to unitary operators (the Heisenberg equation).
The explicit formulae for the semiclassical evolution of the 
Wigner function are given in \cite{RiosOA, OsKon}, whereas
the evolving action of the chord function is presented in \cite{AlmBro06}.
The difficulty that cannot be avoided by changing the representation
lies in the caustics of the initial state, which require more sophisticated
semiclassical treatment. 

A promissing approach lies in the definition
of integral propagators for the Wigner function, or for the chord function.
The former may be defined in terms of the Weyl propagator as a kind of {\it second order
Wigner transform} (see e. g. \cite{RiosOA})
\be
\fl U\!\!\!\!U(\x_t, \x)=
\int {\rm d}\mu\> 
U_{-t}({{\x+\x_t}\over 2}-\mu)\>U_{t}({{\x+\x_t}\over 2}+\mu)
\exp ({2i\over \hbar} \mu \wedge (\x_t-\x).
\ee
Their explicit semiclassical form has been developed in \cite{Dittrich}, 
but these propagators also have their own intrinsic caustics. 
More recently, caustic-free propagators, from the Wigner to the chord function 
and vice versa, have been defined \cite{AlmBro06}. 
These are constructed either in terms of the
propagation of the unitary reflection operators, or the translation operators,
instead of directly evolving the density operator itself.

Several of the geometrical structures underlying the
semiclassical theory of the Wigner function for nonseparable tori
in double phase space that evolve under the action of a general Hamiltonian 
were analyzed in \cite{AlmHan82} for the simplest case where $L=2$.
It will be necessary to push much further this analysis, 
while adapting it to the chord function.
The reason is again that
the partial trace of the density operator is obtained immediately by 
a section through the chord function, $\chi(\vxi_1, \vxi_2=0)$,
which is defined semiclassically by a nonseparable section of 
the double torus that evolved from an initial
product $\tau_\psi \otimes\tau_\psi$. 
For instance, the quantum state and the corresponding double torus
both loose their product form under the action of the simple Hamiltonian employed
in section 8. 
But, slicing through 
a torus produces either a single, or several lower dimensional tori.

This indicates that the semiclassical theory 
of reduced density operators that allow us 
to quantify entanglement, or to calculate the correlations on
separate measurements effected on each component of the system,
can still be associated to lower dimensional Lagrangian surfaces 
that are no longer products.
The problem is to work out the actions and the amplitudes
for this multidimensional geometry. 
This general picture agrees with initial results for nonunitary
Markovian evolution of semiclassical Wigner functions \cite{Alm03}.
It is notable that the same methods which have been used in \cite{Alm82} to show
that the semiclassical approximation to Wigner function satisfy
the purity condition, ${\rm tr}\>\oprho^2=\oprho$, reveal the loss of purity
with time due to decoherence.
 
A final comment about entanglent concerns the choice of components 
which arises with the freedom provided by unitary transformations,
corresponding to classical canonical transformations.
It is possible to {\it disentangle} each state that has lost its original product form
by merely reversing time. This corresponds classically to runing Hamilton's 
equations backwards in the double phase space. 
In the example of section 8, the product of two harmonic oscillator ground states
is recovered from the EPR state.

In quantum mechanics, one can always specify a unitary transformation on the entire
Hilbert space, which transforms any given state into any other state 
and we can choose the latter to be a product state.                          
However, the correspondence
for this disentanglement cannot exist 
for typical quantum eigenstates of a classically chaotic Hamiltonian.
According to Shnirelman's theorem \cite{Shnirelman},
these {\it ergodic states} are supported
by the full energy shell, in the sense that averages of smooth observables
are well approximated by classical averages over this surface.
In the case where $L=2$, the energy shell is 3-D and, 
because it has an extra dimension, there exists no
classical canonical transformation, whether linear or nonlinear, 
that can transform it into the product of two closed curves. 
Therefore, ergodic eigenstates are {\it essentially entangled}
from the point of view of classical correspondence.
As it happens, it is precisely this set of states that has resisted
for decades all attempts at a compact semiclassical characterization.
The study of traces of classical chaos in quantum mechanics is known as
{\it quantum chaology} \cite{BerHouches}. The characterization of ergodic states 
as those that are not classically disentangleable 
creates a bridge between entanglement theory and
quantum chaology. It remains to be seen whether this special type of entangled
state has any application in quantum information theory.

\ack
This text develops a sereis of Lectures delivered at the MPIPKS-Dresden
in September 2005. 
Many of the ideas and results presented here arose
in the context of a fruitful collaboration with Olivier Brodier in CBPF.
Raul Vallejos, Marcos Saraceno and Pedro Rios also
played important roles in the development of the conceptual view
that I have here attempted to sketch.  Maria Carolina Nemes
contributed some helpful criticism of the manuscript and
I also thank the MPIPKS-Dresden for its hospitality during
part of its development.
Partial financial support from 
Millenium Institute of Quantum Information, PROSUL 
and CNPq is gratefully acknowledged.

\end{document}